\documentclass[twocolumn,trackchanges]{aastex62}

\usepackage{hyperref}
\usepackage{graphicx}
\usepackage{soul}
\usepackage{subfigure}
\usepackage{textcomp}
\usepackage{xcolor}

\newcommand{\hi}{\ion{H}{1}}
\newcommand{\hin}{\ion{H}{1}}
\newcommand{\ciii}{\ion{C}{3}}
\newcommand{\oi}{\ion{O}{1}}
\newcommand{\ovi}{\ion{O}{6}}
\newcommand{\ovii}{\ion{O}{7}}
\newcommand{\oviii}{\ion{O}{8}}

\newcommand{\oviiin}{\ion{O}{8}}
\newcommand{\caii}{\ion{Ca}{2}}
\def\hi{{{\rm H}\,\hbox{{\sc i}}~}}
\def\hin{{{\rm H}\,\hbox{{\sc i}}}}
\def\ciii{{{\rm C}\,{\sc iii}~}}
\def\oi{{{\rm O}\,\hbox{{\sc i}}~}}
\def\ovi{{{\rm O}\,{\sc vi}~}}
\def\ovii{{{\rm O}\,{\sc vii}~}}
\def\oviii{{{\rm O}\,{\sc viii}~}}

\def\oviiin{{{\rm O}\,{\sc viii}}}
\def\caii{{{\rm Ca}\,\hbox{{\sc ii}}}}
\def\suzaku{{\it Suzaku}~}
\def\chandra{{\it Chandra}~}

\def\xmmn{{\it XMM-Newton}}

\def\msun{{M$_{\odot}$}}

\begin{document}

\title{\textcolor{black}{Evidence for} massive warm-hot circumgalactic medium around NGC\,3221}
\correspondingauthor{Sanskriti Das} \email{Email: das.244@buckeyemail.osu.edu}
\author[0000-0002-9069-7061]{Sanskriti Das}
\affil{Department of Astronomy, The Ohio State University, 140 West 18th Avenue, Columbus, OH 43210, USA}
\author{Smita Mathur}
\affil{Department of Astronomy, The Ohio State University, 140 West 18th Avenue, Columbus, OH 43210, USA}
\affil{Center for Cosmology and Astroparticle Physics, 191 West Woodruff Avenue, Columbus, OH 43210, USA}
\author{Anjali Gupta}
\affil{Department of Astronomy, The Ohio State University, 140 West 18th Avenue, Columbus, OH 43210, USA}
\affil{Columbus State Community College, 550 E Spring St., Columbus, OH 43210, USA}
\author{Fabrizio Nicastro}
\affiliation{Observatorio Astronomico di Roma - INAF, Via di Frascati 33, 1-00040 Monte Porzio Catone, RM, Italy}
\affiliation{Harvard-Smithsonian Center for Astrophysics, 60 Garden St., MS-04, Cambridge, MA 02138, USA}
\author{Yair Krongold}
\affiliation{Instituto de Astronomia, Universidad Nacional Autonoma de Mexico, 04510 Mexico City, Mexico}
\author{Cody Null}
\affil{Columbus State Community College, 550 E Spring St., Columbus, OH 43210, USA}

\begin{abstract}
\noindent We report a 3.4$\sigma$ detection of the warm-hot, massive, extended circumgalactic medium (CGM) around an L$^\star$ star-forming spiral galaxy NGC\,3221, using deep \suzaku observations. The temperature of the gas is $10^{6.1}$ K, comparable to that of the Milky Way CGM. The spatial extent of the gas is at least $150$ kpc. For a $\beta$-model of density profile with solar abundance, the central emission measure is EM = $3\pm 1 \times 10^{-5}$ cm$^{-6}$ kpc and the central electron density is $n_{eo} = 4\pm 1 \times10^{-4}$ cm$^{-3}$, with a slope of $\beta = 0.56$. We investigate a range of $\beta$ values, and find that the details of the density profile do not change our results significantly. The mass of the warm-hot gas, assuming a metallicity of $\frac{1}{3}$ Z$_\odot$ is $16 \pm 3 \times 10^{10}$ \msun, being the most massive baryon component of \textcolor{black}{NGC\,3221. The baryon fraction is $f_b$ = 0.120 $\pm$ 0.036 (statistical) $^{+0.104}_{-0.048}$ (systematic), consistent with the cosmological mean value, closing the baryon budget of this galaxy.} We also investigated the missing metals problem in conjunction with the missing baryons problem and conclude that metals are likely to be preferentially expelled from the galaxy. Ours is the first detection of an extended warm-hot CGM around an external L$^\star$ star-forming spiral galaxy, where \textcolor{black}{the CGM likely accounts for the missing galactic baryons.} 
%\st{We further investigate the thermodynamics of the warm-hot gaseous halo combining the physical properties of the galactic disk and the CGM.  We find that the CGM can be heated and enriched with metals by the starburst-driven feedback. However, some of the outflowing gas is likely to leave the galaxy, and some is likely to precipitate back onto the disk, providing fuel for the next generation of star-formation.}
\end{abstract}


\section{Introduction} \label{sec:intro}
\begin{table*}
\renewcommand{\thetable}{\arabic{table}}
\centering
\caption{Basic properties of NGC3221 \citep[Taken from][]{Lehmer2010}} \label{tab:galaxy}
\begin{tabular}{c c c c c c}
\tablewidth{0pt}
\hline
\hline
Type & Location & D$_L$ &  Size & log M$_\star$ &  SFR \\
 & ($l,b$) & (Mpc) & & (M$_\odot$) & (M$_\odot$yr$^{-1}$) \\
\hline
\decimals
SBcd & (213.97$^\circ$, 55.71$^\circ$) & 59.46 & 3.2$'\times$0.7$'$ & 11.00$\pm$0.10 & 9.92$\pm$1.00 \\
\hline
%\multicolumn{5}{c}{Obtained }
\end{tabular}
\end{table*}
\noindent It has been known from observations that the nearby galaxies are missing most of their baryons. The stellar and ISM (interstellar medium) components account for a small fraction of the total baryons \citep{Tumlinson2017}, compared to the amount expected from the universal
baryon fraction of $\Omega_b/\Omega_m = 0.157 \pm 0.001$ \citep{Planck2016}. The missing metals problem comes alongside the missing baryons problem; nearby galaxies are short of metals as expected from the star formation history of the universe \citep{Shapley2003}. It is also related to the problem of missing feedback \citep{Wang2010}, where the fate of the star-formation driven outflow remains untraced. A possible solution to all of these problems lies in the highly ionized warm-hot circumgalactic medium (CGM) extended out to the virial radius of the galaxies, as has been predicted by theoretical models \citep{White1978,Ford2014,Oppenheimer2016,Suresh2017}. The CGM of galaxies is supposed to be a large reservoir of warm-hot gas, and can account for $\approx$ 40\% of metals produced by star-forming galaxies \citep{Peeples2014}. This warm-hot (T$= 10^6$-$10^7$K) phase can be
probed by highly ionized metals (e.g. \ovii and \oviiin), the dominant
transitions of which lying in the soft X-ray band. Deep X-ray
observations in emission and absorption are necessary to distinguish
between different sources of the missing baryonic mass and characterize
the medium as a function of the host galaxy properties with a broad
parameter space. The distribution of the density, metallicity and
temperature, the spacial extent, and the mass of this warm-hot gas
provide important constraints to the models of galaxy formation and the
accretion and feedback mechanisms.  

The search for missing mass in the form of warm-hot gas beyond
the optical radii of galaxies started with \textit{ROSAT} and continued
with \textit{Chandra}, \textit{XMM-Newton} and \textit{Suzaku}. However,
unlike the rich galaxy clusters \citep{White1993} and the massive
early-type galaxies \citep{Forman1985,Humphrey2012}, where ample amount of hot
gas dominates the baryonic component of the system and retains the
cosmological allotment of baryons, the X-ray coronae around spirals are
faint, which makes their detection challenging. While the warm-hot CGM detected around the Milky
Way may account for the missing mass
\citep{Gupta2012,Gupta2014,Nicastro2016b,Gupta2017},
the extended CGM in X-ray emission has been confidently detected only
around massive galaxies ($M_\star>2\times10^{11} M_\odot$), and only out
to a fraction of their virial radii, with mass insufficient to close
their baryonic budget
\citep{Anderson2011,Dai2012,Bogdan2013a,Bogdan2013b,Anderson2016,Bogdan2017,Li2017,Li2018}.

In this paper, we search for hot diffuse gas around an L$^\star$ spiral
galaxy NGC\,3221. The basic properties of the galaxy are
given in Table \ref{tab:galaxy}. It is an actively star-forming galaxy
with a high star-formation rate. It has a high ratio of
$L_{FIR}/D_{25}^2 = 13.8 \times 10^{40}$ erg s$^{-1}$kpc$^{-2}$ where
$L_{FIR}$ is the far-infrared luminosity and $D_{25}$ is the diameter
out to the surface brightness of $25$ mag arcsec$^{-2}$.  Additionally,
it has a high ratio of flux densities at 60 and 100 microns,
$S_{60}/S_{100} =0.37$, confirming that NGC\,3221 is
an actively star-forming galaxy \citep{Rossa2003}. 

 Our paper is structured as follows: we discuss the data reduction and analysis in section \ref{sec:red} and \ref{sec:anl}, starting with data reduction followed by the point source identification, imaging analysis and spectral analysis of the diffuse medium from the \textit{Suzaku} data. Then we report the detection of the CGM emission, model its radial profile and derive some of its physical properties in section \ref{sec:results}. Our findings are interpreted in the context of
missing baryons, missing metals and missing feedback problems in section \ref{sec:discuss}. Then, we compare our result with earlier observations in section \ref{sec:compare}. We summarize our results and outline some of
the future aims in the last section. 

\section{Observations and data reduction} \label{sec:red}
%----------------------------
\subsection{\textit{Suzaku} Observations} 
%-----------------------------
\noindent We observed the field of NGC\,3221 with \textit{Suzaku}. 
As noted above, the emission signal from the CGM is expected to be faint, dominated by foreground/background. To characterize the foreground and background, we also observed two fields
$\approx 2\deg$ away from the galaxy. From now on, we will refer to the
field of the galaxy as the galaxy-field and the other two fields as
off-fields. With the deep \textit{Suzaku} observations, our goal is to
extract the emission signal from the CGM of NGC\,3221, and the
off-fields are used to determine the foreground/background emission. We
expect to detect emission from about a million degree thermal plasma,
with the most dominant signature being the emission lines of \ovii
and/or \oviii around 0.5--0.6 keV. Therefore, the soft X-ray band is
important for our analysis, as discussed below. 

 The large field of view (FOV $\approx 17.8'\times 17.8'$), low and
stable detector background, and high sensitivity to detect low surface
brightness in soft X-rays have made \textit{Suzaku} an excellent choice
to study the diffuse circumgalactic medium. The back-illuminated X-ray
Imaging Spectrometer (XIS-1) with largest effective area among all other
chips at our energy range of interest (0.4-5.0 keV) serves the best for
this purpose. We observed the NGC\,3221 galaxy-field and off-field2 in
November, 2014 and off-field1 in May, 2014.  The unfiltered exposure
time of the galaxy-field and the off-fields are $\approx$121 ks, 41 ks
and 40 ks respectively.
%---------------------------------------------------------
\subsection{Data Reduction}\label{sec:data-process}
%----------------------------------------------------------
\noindent We reduced the data very carefully taking into account the
changes in XIS-1 instrumentation with time and also the effect of
enhanced solar activity on the post-2011 observations of a low
earth-orbit ($\approx 550$ km) satellite like \suzaku (Appendix
\ref{data-reduction}).
\begin{figure*}
\begin{subfigure}
\centering
\includegraphics[trim=50 0 50 0,clip,scale=0.3]{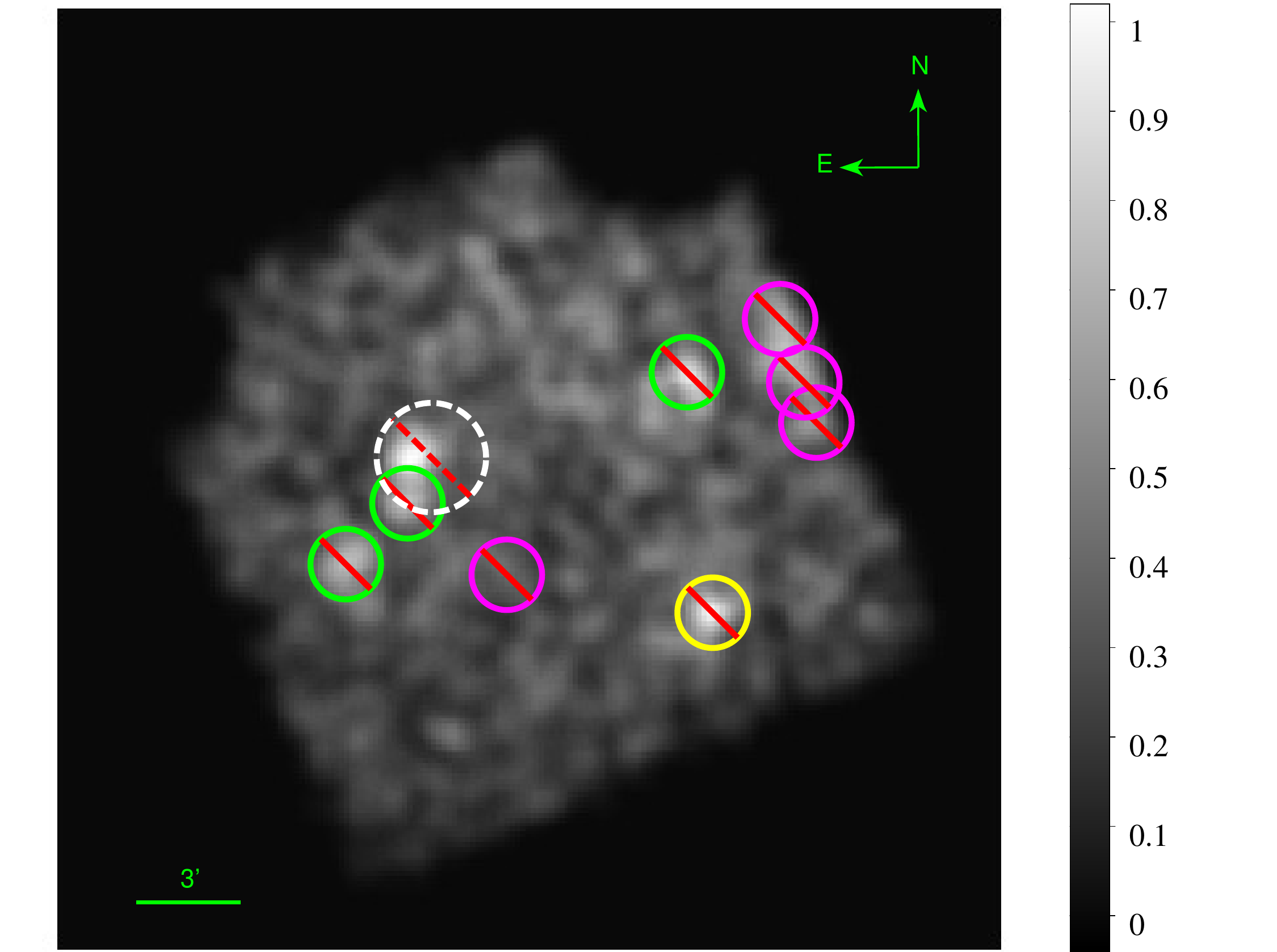}
\end{subfigure}
\begin{subfigure}
\centering
\includegraphics[trim=50 0 50 0,clip,scale=0.3]{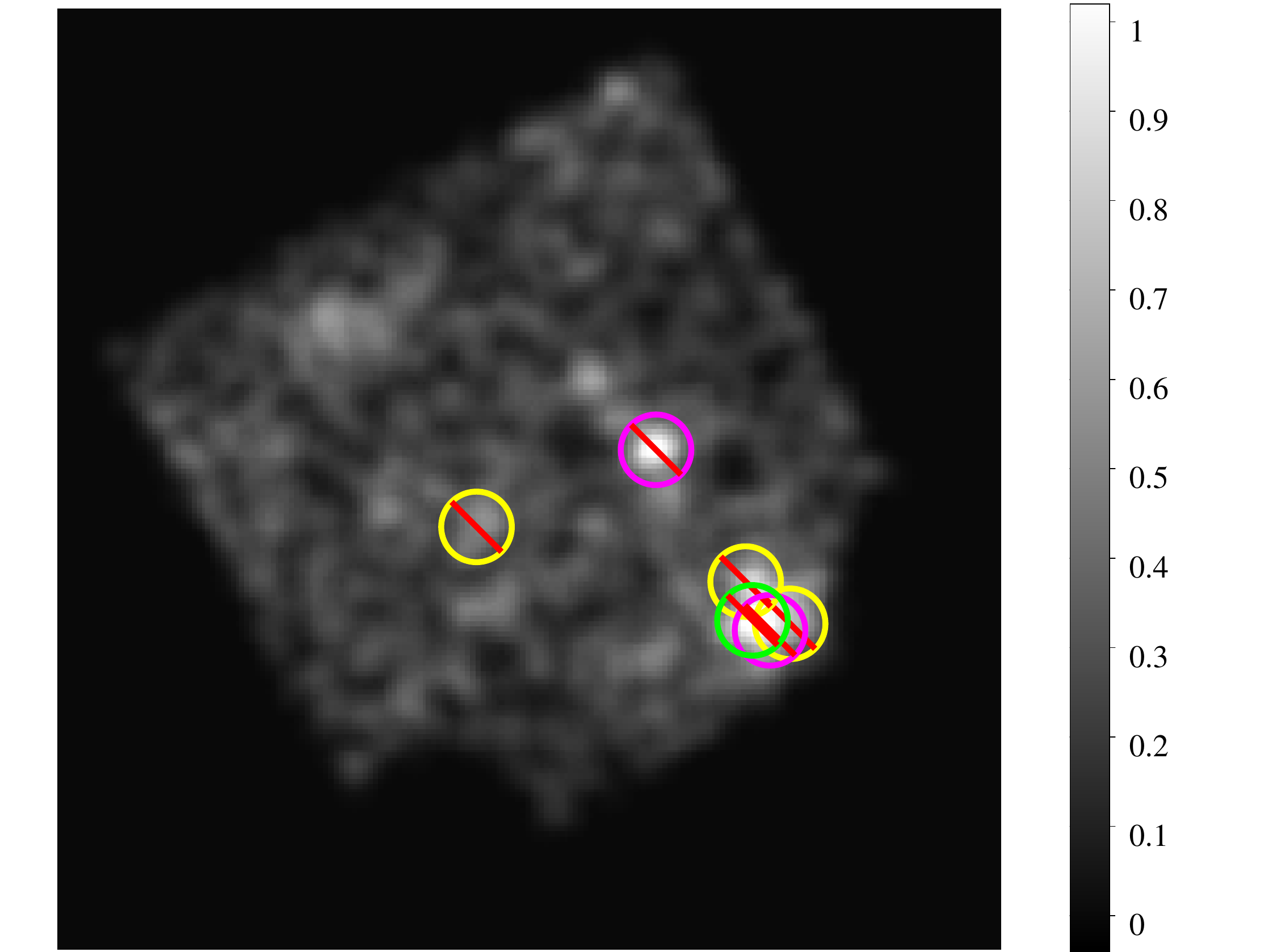}
\end{subfigure}
\begin{subfigure}
\centering
\includegraphics[trim=50 0 55 0,clip,scale=0.3]{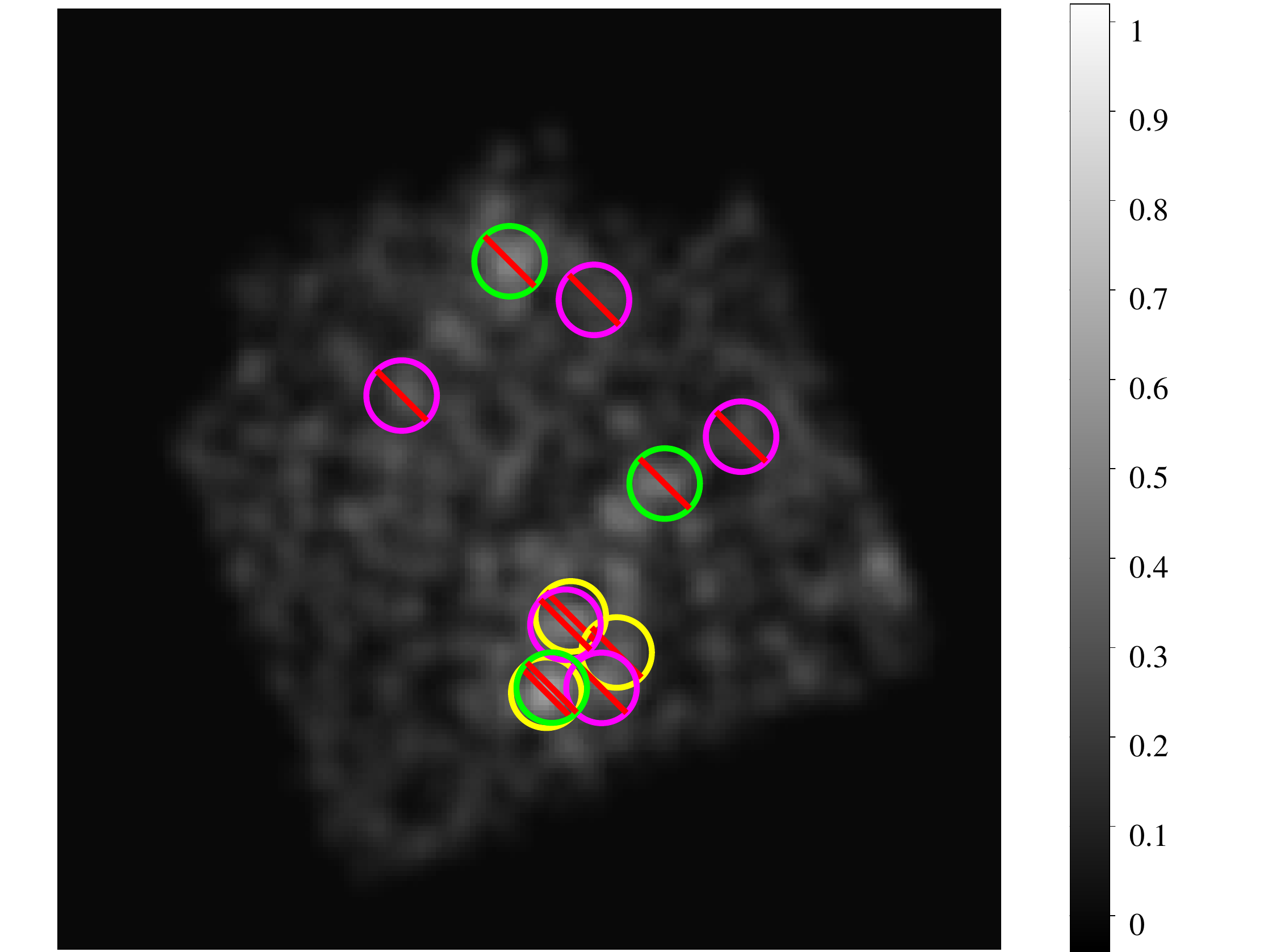}
\end{subfigure}
\caption{0.4-2.0 keV images of the \textit{Suzaku} fields (left:
  galaxy-field, middle \& right: off-fields). Point sources identified
  in 0.4-0.7 keV, 0.7-1.0 keV and 1.0-2.0 keV are shown in magenta,
  yellow and green respectively, smoothed to the PSF. The white dashed
  circle in the galaxy-field covers the optical extent of NGC\,3221. The
  sky direction and the angular scale (16 kpc$\approx$1$'$) are shown in
  the left panel \label{fig:soft_src}}
\end{figure*}

Our first task was to identify and remove the point sources in the three
observed fields. We identify the compact bright sources in 0.4-0.7
keV, 0.7-1.0 keV and 1.0-2.0 keV bands separately.  We smoothed the images with Gaussian kernel radius of 5 to identify sources $\geqslant 3\sigma$ brighter than their background (figure \ref{fig:soft_src}). We smoothed all sources upto the spatial resolution of \textcolor{black}{\suzaku (PSF $\approx$ 1.8$'$--2.0$'$, half power diameter, or radius of 0.9$'$--1.0$'$)} unless the source itself appeared larger, and removed the contribution of all sources from the respective fields. Point source contamination in the hard ($>$2 keV) band is modeled spectroscopically, as discussed below in \S \ref{sec:image_analysis}. The projected semi-major axis of the stellar disk of NGC\,3221 is $1.6' \approx 25$ kpc. To separate the large scale diffuse X-ray emission of the CGM from the emission of the stellar disk and the extra-planar region \textcolor{black}{between 5.5 kpc (the semi-minor axis of NGC\,3221) and 25 kpc}, we remove a circular region of 25 kpc radius around the center of NGC\,3221 in the galaxy-field. \textcolor{black}{The excluded region is significantly larger than the \textit{Suzaku} PSF, therefore we do not expect the emission from the stellar disk of NGC\,3221 to contaminate the emission from the extended CGM beyond the radii of $1.6'$}. We construct the count-rate histogram of events observed in 0.4-2.0 and 2.0-5.0 keV bands. We find that there are $\approx3\%$ events outside 2$\sigma$ limit of mean count rate in all the three fields in
both the energy ranges, well within the distribution of Poisson
fluctuations. 
%-----------------------------------------------------
\section{Analysis}\label{sec:anl}
%--------------------------------------------------------
\noindent \textcolor{black}{We have done the data analysis in two steps: imaging and spectral analysis. Before we delve into spectral analysis to extract the CGM signal, we study the radial and angular profile of the soft X-ray imaging data. We clarify that this data is not foreground/background-subtracted. However, as the foreground/background is not spatially correlated with the target galaxy, they are unlikely to show any radial/azimuthal variation with respect to the position of NGC\,3221.}
%----------------------------------------------------
\subsection{Imaging Analysis}\label{sec:image_analysis}
\begin{figure}[h]
\centering
\includegraphics[trim= 650 10 650 20, clip, scale=0.4]{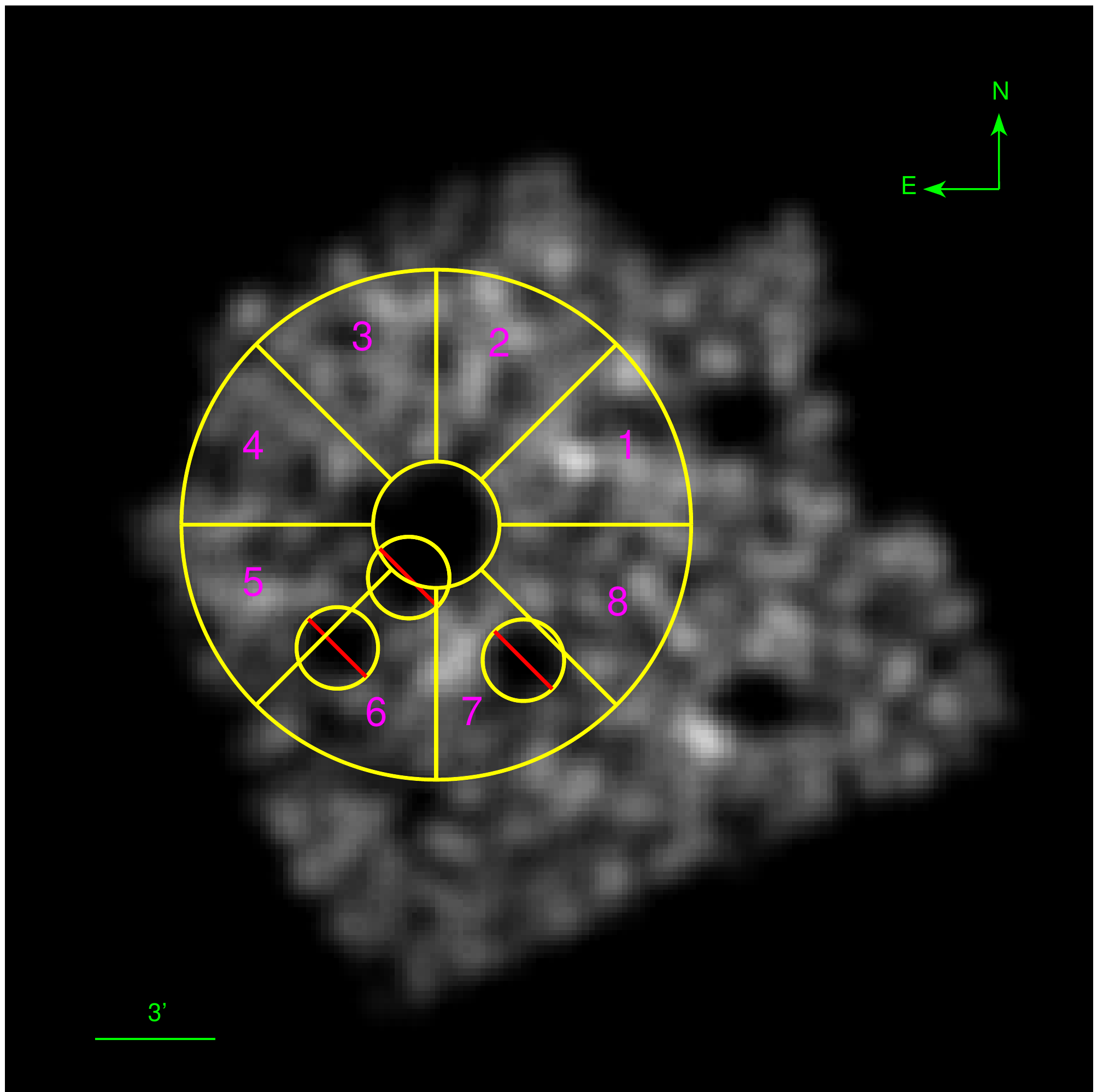}
\caption{Wedges to study the azimuthal surface brightness profile. The inner and outer radial boundaries of the wedges correspond to 25 kpc and 100 kpc from the galaxy. The sky direction and angular scale are shown in the top-right and bottom-left corners respectively \label{fig:wedges}}
\end{figure}

\begin{figure}
\centering
\includegraphics[trim=30 0 00 38, clip, scale=0.475]{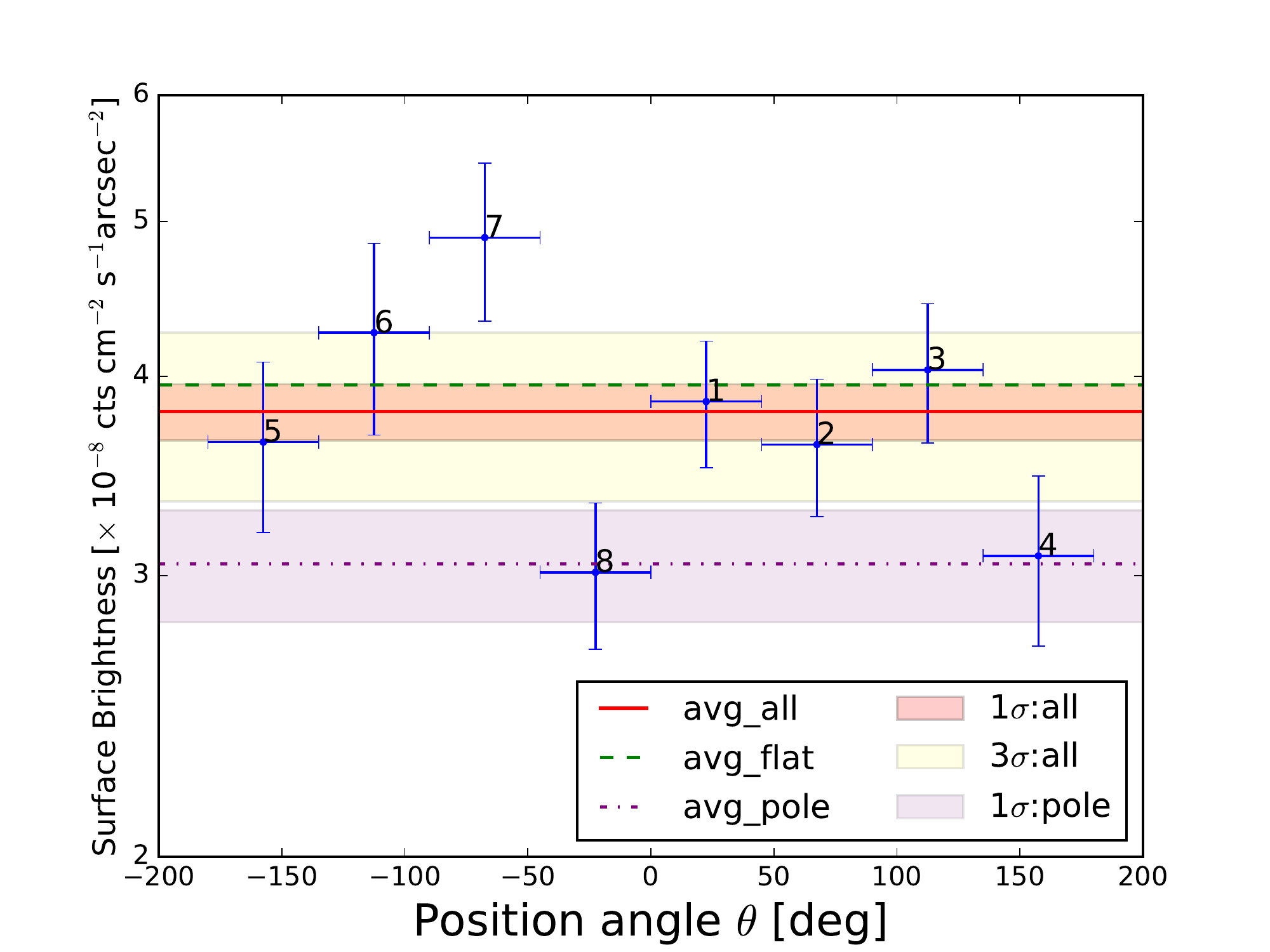}
\caption{Azimuthal surface brightness profile in 0.4-1.0 keV, with the
  wedges labeled as in figure \ref{fig:wedges}. The global mean (avg\_all)
  is shown with the red horizontal line and 1$\sigma$ and 3$\sigma$
  regions are shown with orange and yellow regions,
  respectively. Sectors 4 and 8 are along the minor axis of the galaxy,
  with significantly lower surface brightness; their average (avg\_pole;
  dash-dotted line) and the 1$\sigma$ region are shown in purple. Sector 7
  and 8 are 3$\sigma$ away from the global average. The average of
  sectors 7 and 8 (avg\_flat) is shown with the dashed green line. As
  discussed in the text, these azimuthal variations do not affect the
  determination of the radial profile.}\label{fig:surf-az}
\end{figure}

%----------------------------------------------------
\noindent 
The largest complete annular region we can extract is between 25 kpc and
100 kpc. To study the azimuthal variation of emission, we split this annular
region into 8 wedges each with an opening angle of 45$^o$
(figure \ref{fig:wedges}).  
The surface brightness in 0.4-1.0 keV is shown
in figure \ref{fig:surf-az}, with a data point for each wedge. The red
line shows the average surface brightness ($3.80\pm0.15 \times$
10$^{-8}$ counts cm$^{-2}$ s$^{-1}$arcsec$^{-2}$), with the shaded
region showing $1\sigma$ error. The surface brightness in sectors 1, 2,
3, 5 and 6 agrees with each other within statistical uncertainties,
indicating that the extended emission is fairly uniform. There is an
apparent increase of the surface brightness in sector 7; we interpret
this as due to the contamination from a soft compact source leaked
beyond the half-power diameter of 2$'$ (see figure \ref{fig:soft_src}, top
panel). The sectors 4 and 8 have a mean surface brightness = $3.05\pm
0.25 \times 10^{-8}$ counts cm$^{-2}$ s$^{-1}$arcsec$^{-2}$, shown by
the dash-dotted line, with the purple shaded region showing $1\sigma$
uncertainty. This is $\approx$40\% smaller than the global average, but
is only $2\sigma$ from the mean, so we do not draw any strong inference
from this deviation. We note, however, that these sectors are along the
minor axis of NGC\,3221, so the low surface brightness could be the
manifestation of a cavity created by the bipolar outflow from this
star-forming galaxy. Such a cavity has been observed in the Milky Way
\citep{Nicastro2016a}. Overall, the uniform distribution is
morphologically consistent with the quasi-static warm/hot gas residing in the potential
well of the galaxy. 

\begin{figure}
\centering
\includegraphics[trim= 700 0 700 30, clip, scale=0.5]{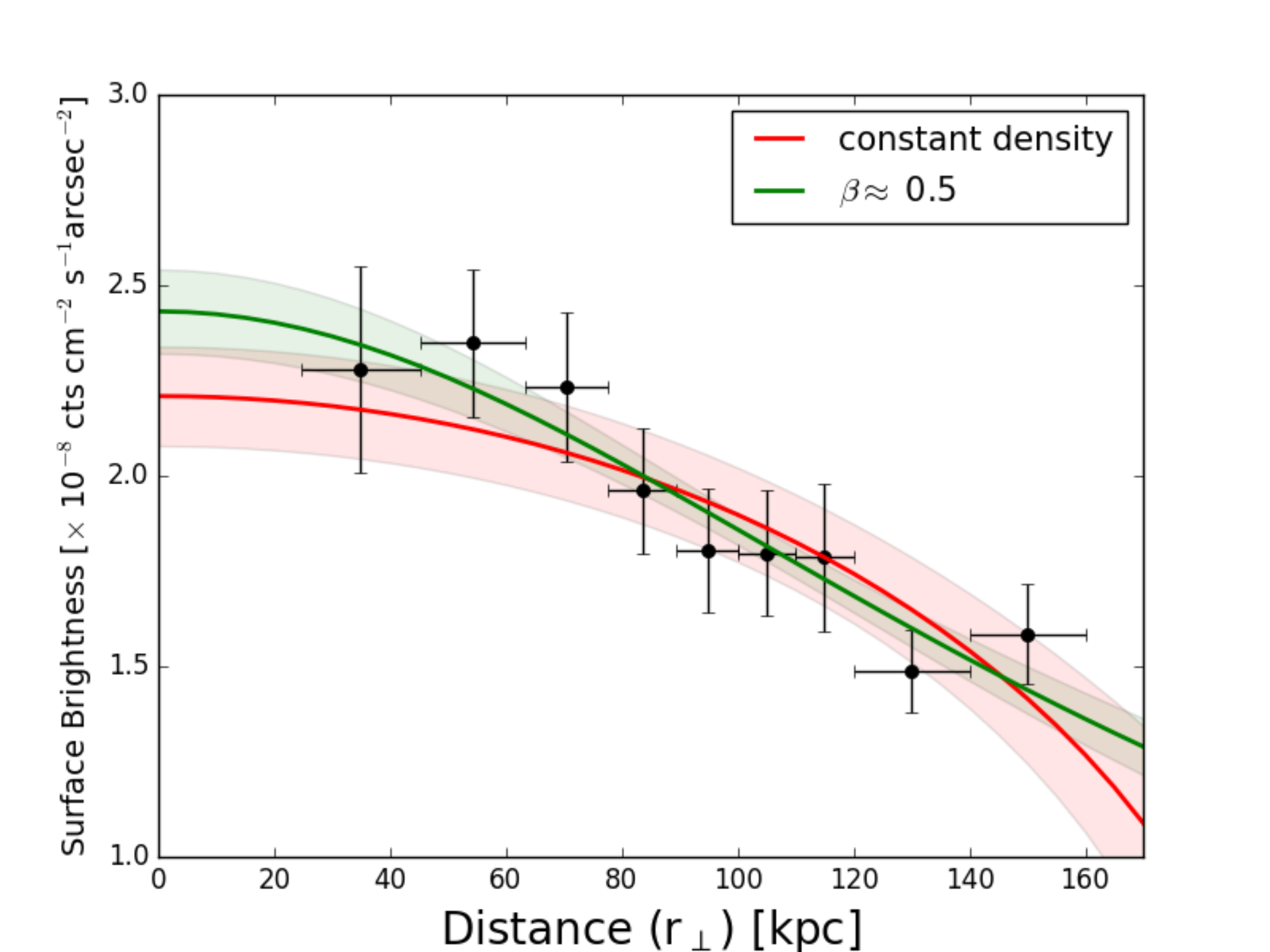}
\caption{Radial surface brightness profile of NGC\,3221 in 0.4-1.0
  keV. The shaded regions represent 68\% confidence interval,
  considering the uncertainties in the best-fitted values of each
  model's free parameters \label{fig:surf-rad}}
\end{figure}

To study the radial variation of the surface brightness in 0.4--1.0 keV,
we extract 9 annular regions of similar geometric area (i.e. (r$_{out}^2$-r$_{in}^2$) is constant) between 25 kpc and 160 kpc. The azimuthal
variations discussed above are averaged out. The surface brightness
remains almost flat near the core and then slowly decreases with
distance from the galaxy (figure \ref{fig:surf-rad}). \textcolor{black}{Although the surface brightness is dominated by the foreground/background, its contamination over the FOV in the galaxy-field should be constant. The transverse radial variation of the surface brightness with respect to the galaxy's position indicates the existence of a medium related to NGC\,3221. Otherwise, a flat radial surface brightness profile would indicate that the halo emission from NGC\,3221 is too faint to be detected, which is not the case. Therefore, }we fit the surface brightness profile for two density distributions, with a truncated constant density medium and with a $\beta$-model. For the homogeneous (constant density) medium, we use equation \ref{eq:S_homo}:
\begin{equation}\label{eq:S_homo}
S(r) = S_o\sqrt{1-(r_\perp/R_{out})^2}
\end{equation}
where R$_{out}$ is the spatial extent of the gaseous medium. For the $\beta$-model we use equation \ref{eq:S_beta} \citep{Sarazin1986}:
\begin{equation}\label{eq:S_beta}
S(r) = S_o (1 + (r_\perp/r_c)^2 )^{-3\beta+0.5}
\end{equation}
where S$_o$ is the central surface brightness, r$_c$ is the core radius
and $r_\perp$ is the projected distance across the line of sight. The
best-fit values ($\chi^2/dof = 7.07/7$) for the constant density model
are : S$_o$ = $(2.2\pm 0.1)\times10^{-8}$ counts cm$^{-2}$
s$^{-1}$arcsec$^{-2}$ and R$_{out}$= 195 $\pm$ 14 kpc. We fit the
$\beta$-model as follows. We first fix $\beta=0.5$ as is usually done in literature (see
\cite{Gupta2017} and references therein) and fit the radial profile
for S$_o$ and r$_c$. The best-fit value of r$_c$ is $178 \pm 17$
kpc. Then we fix r$_c$ at 178 kpc and fit for S$_o$ and $\beta$. The
resulting best-fit values ($\chi^2/dof = 3.76/7$) of the parameters are:
S$_o$ = ($2.4\pm 0.1)\times$10$^{-8}$ counts cm$^{-2}$
s$^{-1}$arcsec$^{-2}$, and $\beta= 0.50\pm 0.05$. 
\textcolor{black}{The central surface brightness is the aggregated emission from the foreground, background and the halo emission, while the radial parameters (R$_{out}$, r$_c$ and $\beta$) are the characteristics of the density profile of the NGC\,3221 halo.} This is further confirmed with the spectral analysis below. 
%------------------------------------------------------
\subsection{Spectral Analysis}\label{sec:spec_analysis}
%------------------------------------------------------
\noindent For the spectral analysis we generate the non-X-ray background (NXB) and
the redistribution matrix function (RMF) for each field (see the details
in appendix \ref{data-reduction}). We then rebin each spectrum such
that there is no bin with zero variance, and any spectral information is
not lost due to over-smoothing. As the effective area of XIS is very
small below 0.4 keV and the detector background is quite high above 5.0
keV, we concentrate on the energy range of 0.4-5.0 keV, for
spectral analysis. 

The spectra of the diffuse background are complicated, including
multiple components which are spectroscopically resolved \citep{Henley2010,Henley2013,Gupta2014,Henley2015a,Henley2015b,Gupta2017}. Our
goal is to detect the CGM around NGC\,3221, but the galaxy-field spectrum
contains the CGM emission plus all the background and foreground
emission present in the off-fields. Therefore, we first fit the off-fields
spectra and then use those models in fitting the galaxy-field
spectrum. Accordingly, we fit the off-field spectra as a composite of
three components:\\
1) \textcolor{black}{The foreground consists of the local hot bubble (LHB) and solar wind charge exchange (SWCX). As the target is at a relatively low ecliptic latitude ($\beta$ = 10.62$^\circ$), the heliospheric SWCX is likely present in soft X-ray. But, it does not appear to contribute significantly to the Oxygen lines of interest \citep{Miller2015}.} The components of LHB and SWCX cannot be separated at the spectral resolution (FWHM $\approx$ 0.05 keV at 1 keV) of XIS \citep{Yoshino2009}. So, the combined emission from the LHB and SWCX is modeled as an unabsorbed collisionally-ionized plasma in thermal equilibrium.  We fix the temperature of this component at $k_BT = 0.099$ keV and the metallicity at solar \citep[see][]{Gupta2014,Henley2015b,Gupta2017}.\\
2) Collisionally ionized plasma in thermal equilibrium, representing
  the warm-hot gaseous halo of the Milky Way (MWH), absorbed by the
  Galactic ISM. Again, we fix the metallicity at
  solar. The normalization factor of the thermal plasma model is
  metallicity-weighted, so the exact value of the input metallicity does
  not matter. Our aim is to merely include the contribution of the MWH
  in the spectral analysis and keeping Z = Z$_\odot$ does not affect 
  the final result.\\
3) Absorbed power law to account for the unresolved extragalactic point sources,
  forming the cosmic X-ray background (CXB). We keep the normalization
  and the power law index as a free parameter.

We model the thermal plasma using Astrophysical Plasma Emission Code
(\texttt{apec}), which predicts the emission spectrum of optically thin diffuse
gas in collisional ionization equilibrium using the atomic
data from the Astrophysical Plasma Emission Database (APED)\footnote{The atomic database
  (\url{http://www.atomdb.org/physics.php}) includes 
  APED and the spectral models output from
  the \texttt{apec}. The APED files contain information such as wavelengths,
  radiative transition rates, and electron collisional excitation rate
  coefficients. \texttt{apec} uses these data to calculate spectra. \texttt{apec} outputs
  separate continuum and line emissivity files, making it easy to model
  continuum and line emission separately as well as together}. We obtain
the Galactic column density values, N(\hin), toward our fields from the
general tools of HEASARC. We also take into account thermal line
broadening while fitting the spectra. 

\begin{figure*}
\begin{subfigure}
\centering
\includegraphics[trim=10 12 5 10,clip,scale=0.45]{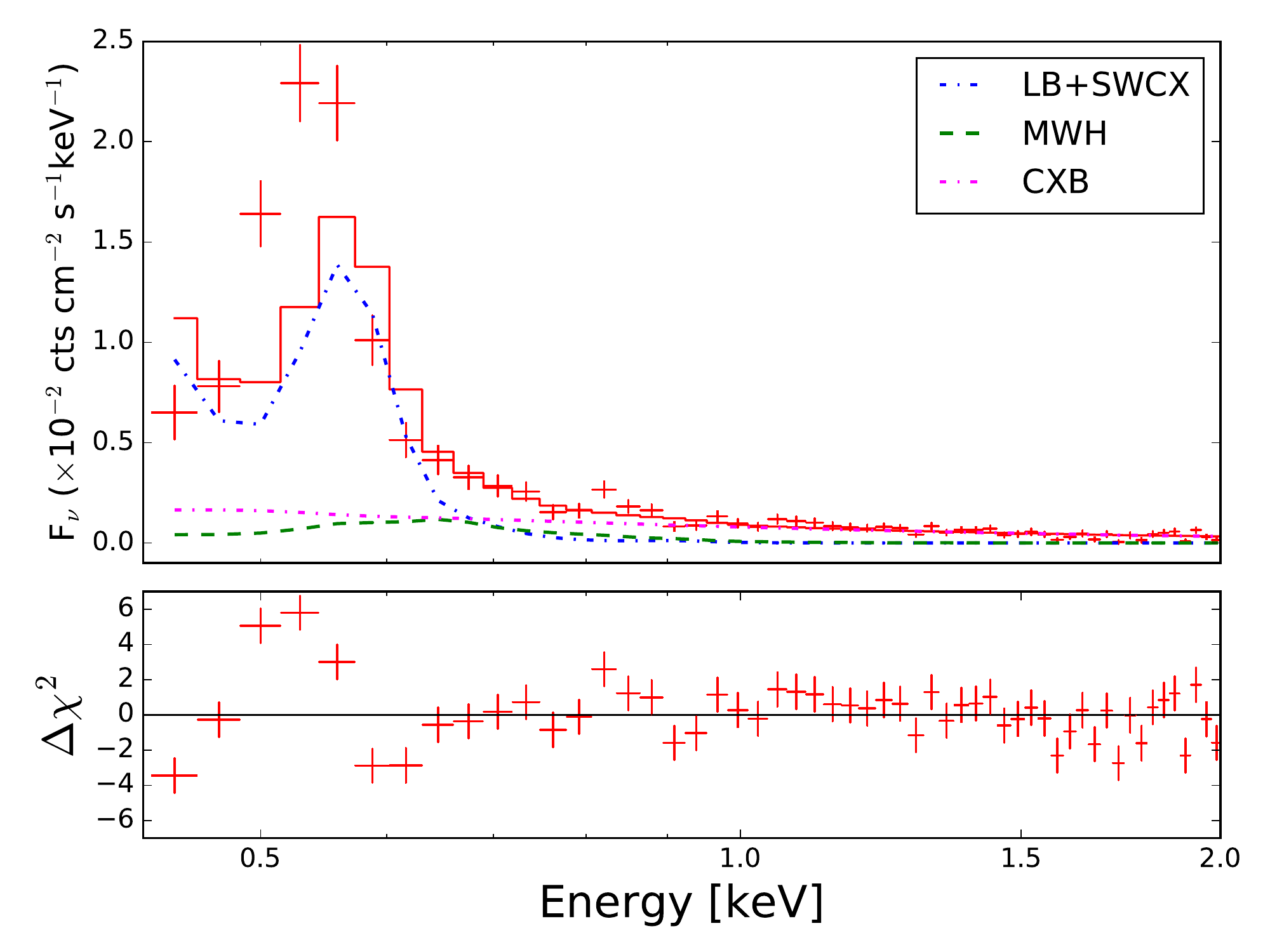}
\end{subfigure}
\begin{subfigure}
\centering
\includegraphics[trim=10 12 5 10,clip,scale=0.45]{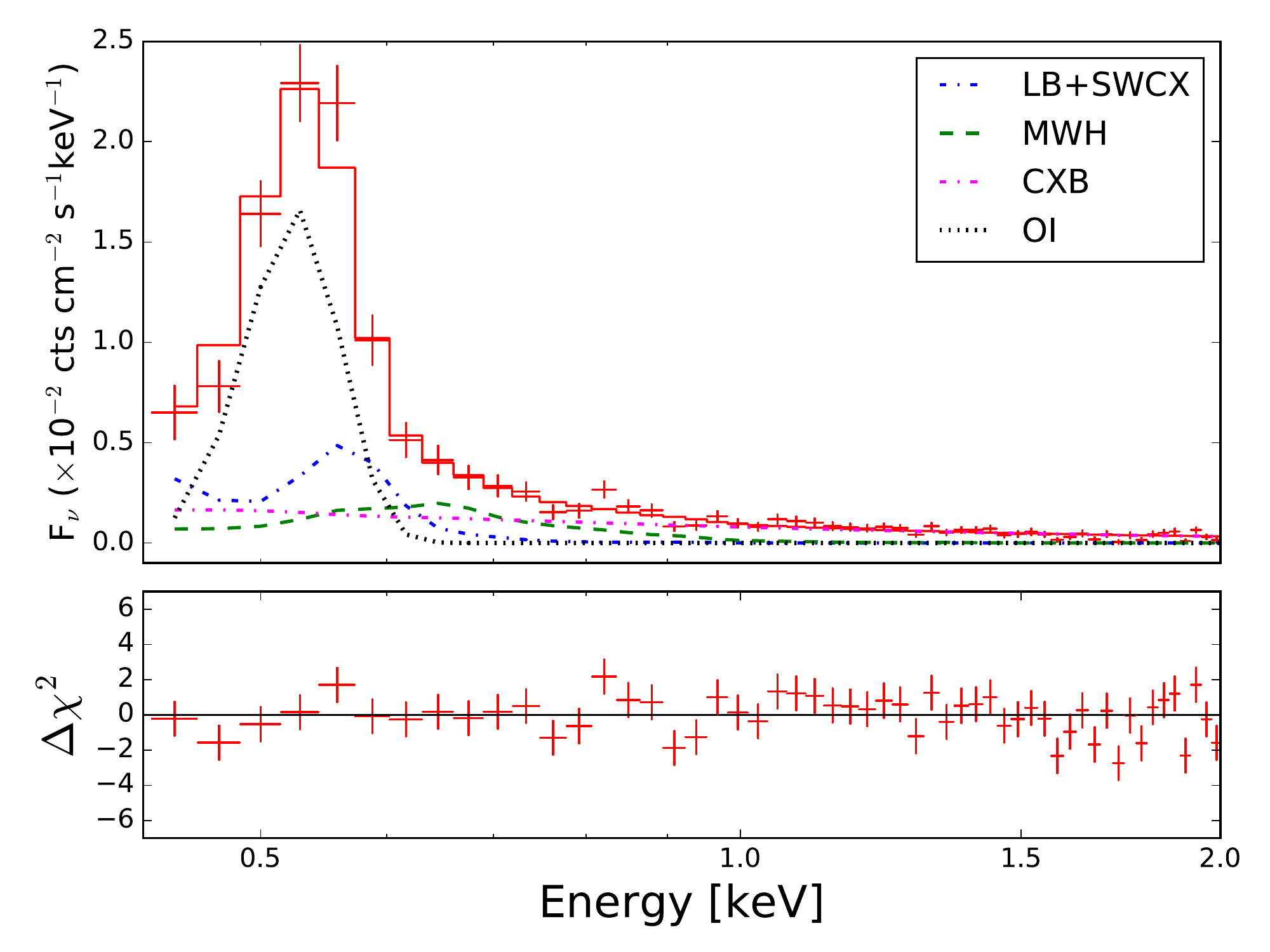}
\end{subfigure}
\caption{Soft spectrum of off-field 2 (covering the whole FOV) and the
  best-fitted models without (left) and with (right) the \oi line. In
  the absence of \oi line the contribution of LHB+SWCX is overestimated
  (top) to adjust the excess due to \oi \label{fig:OI} }
\end{figure*}

In the top panel of figure \ref{fig:OI}, we show the off-field2 spectrum
with the best fitted model containing the three components noted above. A
significant excess in the data around 0.5 keV is clearly seen, leading to
a very poor fit ($\chi^2/dof$ = 195.91/105) and poorly constrained parameter values. This could not be adjusted either by varying the temperature of the LHB+SWCX, or by fixing the temperature of the MWH or allowing the metallicity to vary; this shows
that the excess is not related to either of the plasma models. We
identify the excess as the contamination by \oi fluorescent line at
0.525 keV which is created by fluorescence of solar X-rays with neutral
Oxygen in the Earth's atmosphere, discussed further below. 

\citet{Sekiya2014} pointed out that due to the increased solar
activity the flux of \oi line at $\approx$0.525 keV has increased with
time since 2011 in the \textit{Suzaku}/XIS observations. Since XIS
cannot distinguish the \oi line from the \ovii K$\alpha$ triplet line
(forbidden, inter-combination and resonance lines at 0.561, 0.569 and
0.574 keV respectively) owing to its energy resolution of $\approx 0.05$
keV, unless the \oi fluorescent line is taken into account, the \ovii
line intensity would be significantly overestimated. As shown in the top panel of figure \ref{fig:OI}, we confirm the findings of \citet{Sekiya2014},
now with a more robust spectral modeling (including LHB+SWCX, MWH and
CXB components) and in a larger wavelength window. This reinforces the
importance of considering \oi line contamination in the post-2011 data. 

From here onward, we include in the spectral model an unabsorbed Gaussian emission line fixed at 0.525 keV for the \oi fluorescent line. \textcolor{black}{So, the off-fields now have 4 spectral components: \oi line, LHB+SWCX, MWH and CXB.} We fit the off-field spectra again and and obtain a good fit ($\chi^2/dof$ = 102.68/105), as shown in the bottom panel of figure \ref{fig:OI}. The best fit model of the
off-field defines the background plus foreground model for the galaxy
field.  

Next, we simultaneously fit the galaxy-field between 25 kpc and 200 kpc of NGC\,3221, and the off-fields
spectra. We use the best-fit models of the off-fields as initial guesses. \textcolor{black}{While the off-fields have 4 spectral components, the galaxy-field has 5 spectral components. In the galaxy-field, we add an \texttt{apec} component at the same redshift of NGC\,3221, absorbed by the Galactic ISM, representing the CGM of NGC\,3221. The free parameters are the temperature and metallicity-weighted normalization factor. The \oi line intensity depends on the solar activity during the observations, so we do not tie it in the galaxy-field and the off-fields. As the fields are separated by only 2$\deg$,} we do not expect any significant change in the LHB+SWCX component between the off-fields and the galaxy-field. So, we constrain this component to be the same in the galaxy-field and the off-fields. It is known from previous measurements \citep{Yoshino2009, Henley2013, Gupta2014, Henley2015b, Gupta2017} that the temperature of the MWH remains almost constant while the emission measure can vary by more than an order of magnitude along different sightlines. Therefore, we force the temperature of MWH in the off-fields and the galaxy-field to be the same and let the normalization factor vary within the observed range \citep{Henley2013,Henley2015b}. The CXB is allowed to be the different in the galaxy-field and the off-fields. Tying it to be same did not produce any appreciable change in the fit.
%-----------------------------------------------
\section{Results}\label{sec:results}
\subsection{Detection of the signal}\label{sec:detect}
\begin{figure*}
\begin{subfigure}
\centering
\includegraphics[trim=10 12 5 10,clip,scale=0.45]{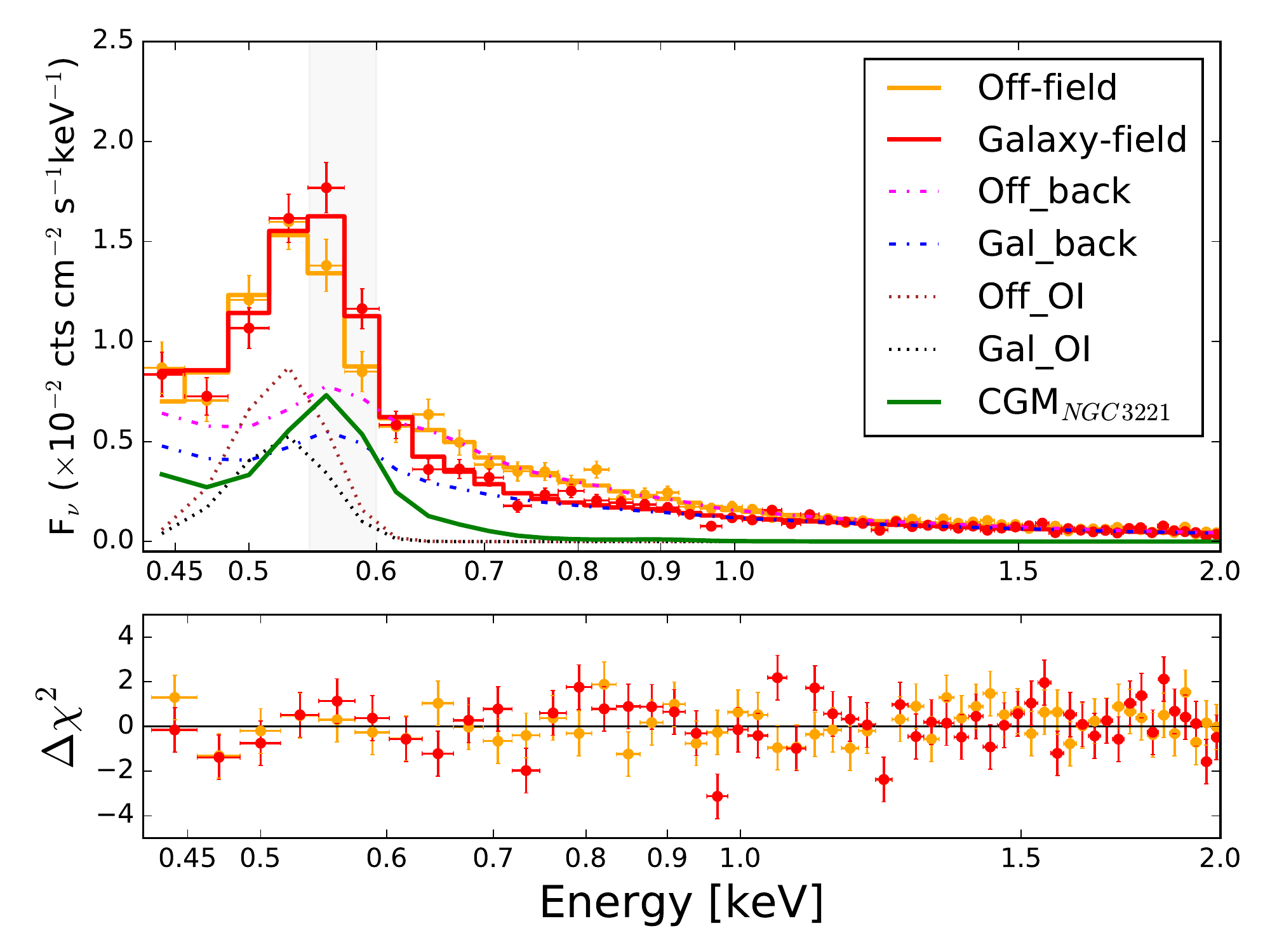}
\end{subfigure}
\begin{subfigure}
\centering
\includegraphics[trim=10 12 5 10,clip,scale=0.45]{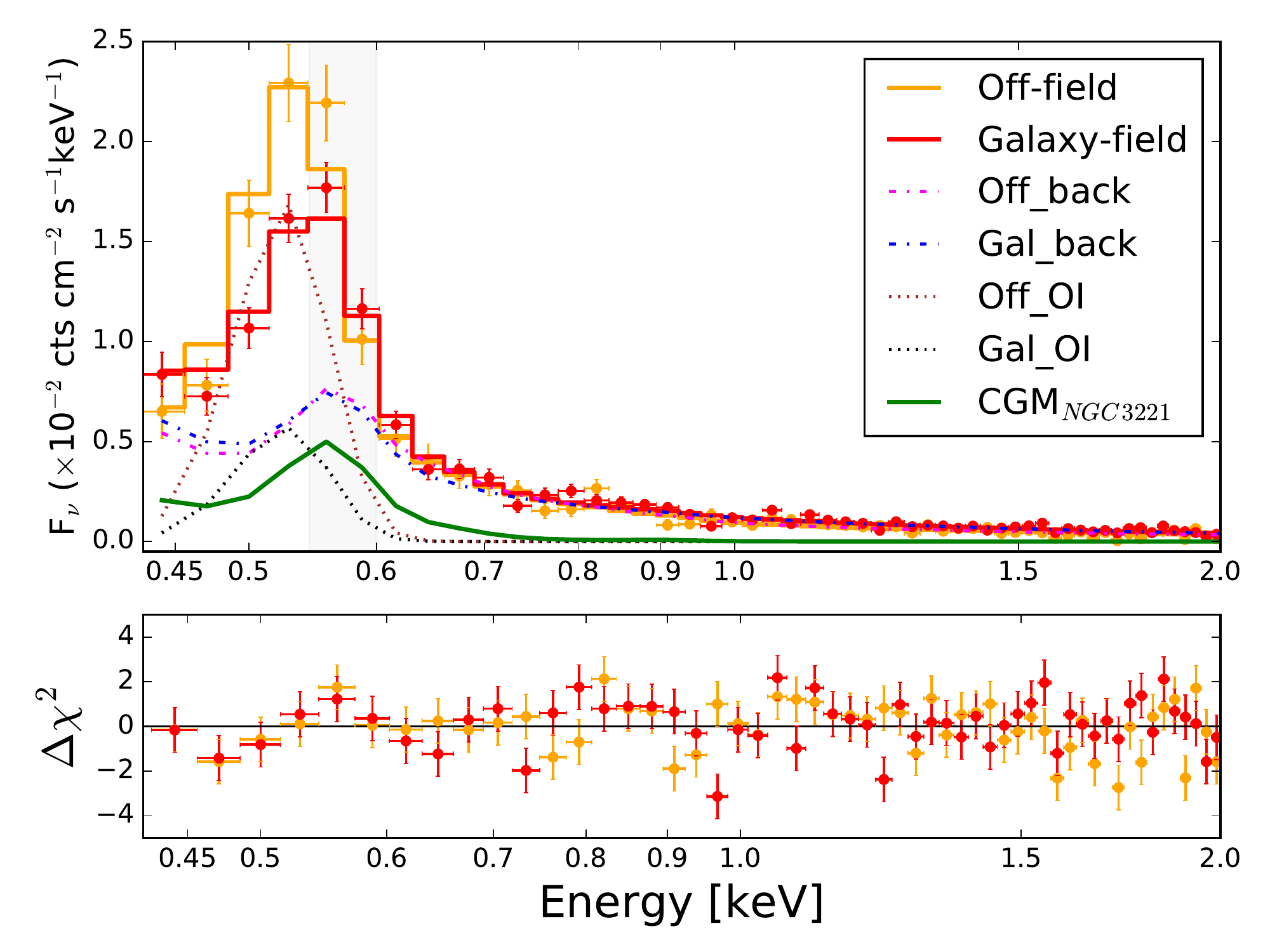}
\end{subfigure}
\caption{Background subtracted spectra of the whole \textit{Suzaku}
  fields, galaxy-field extending out to $\approx$ 200 kpc from the
  center of NGC\,3221, (left: galaxy-field and the off-field 1, right:
  galaxy-field and the off-field 2) with the best-fitted models. The combined emission from the foreground and the background (LHB+SWCX+CXB) has been denoted as $back$ in the legend. An
  excess around 0.575 keV in the galaxy-field with respect to the
  off-field is prominent in the left panel, indicating the existence of
  the CGM signal. Because of larger \oi contamination in the off-field2,
  the wing of the \oi line centered at 0.525 keV overlaps with the \ovii
  line at 0.575 keV (right panel), making the presence of the CGM
  signal not visually evident in the spectrum } \label{fig:spectral-fit}
\end{figure*}

%-------------------------------------------------
\noindent With the spectroscopic modeling discussed above, we isolate the signal
from the CGM of NGC\,3221. Against the background/foreground defined by
off-field1, the emission integral of the galaxy CGM is found to be
$1.1^{+0.4}_{-0.3}$ cm$^{-6}$kpc$^3$, a $3.4\sigma$ detection (note that
the observed emission integral is degenerate with metallicity; the
numbers quoted here are for solar metallicity). Against off-field2, we
detect the CGM signal at $\approx2\sigma$ significance, with
EI$=0.6^{+0.3}_{-0.4}$ cm$^{-6}$kpc$^3$. On average, the halo emission
integral in the whole field is $\approx0.86 \pm 0.24$ cm$^{-6}$kpc$^3$,
showing a significance of 3.6$\sigma$. The measured temperature against off-field1 and off-field2 are consistent with each other: log (T/K) = $6.1^{+0.3}_{-0.2}$ against off-field1 and $6.1^{+0.4}_{-0.2}$ against off-field2 (90\% confidence interval). We find that while the \oi contamination in the off-field1 and the galaxy-field are similar within a factor of $\approx$1.5, it is larger by a factor of $\approx$3 in the off-field2 (figure \ref{fig:spectral-fit}).  As the position and characteristics of the active regions of the solar corona differ by day, it is conceivable that the flux of solar X-ray and the resulting \oi fluorescent line are different in these fields, which are not observed at the same time. Therefore, in further discussions, we use the results of the fit using only the off-field1 as the reference; it gives us better constrained and consistent parameter values. 
\begin{table}
\renewcommand{\thetable}{\arabic{table}}
\centering
\caption{Best-fit values \footnote{Obtained for the whole field, extended out to $\approx$200 kpc around the disk of NGC\,3221} of the CGM component} \label{tab:spectral-fit}
\begin{tabular}{c c c c}
\tablewidth{0pt}
\hline
\hline
Reference & kT [keV] & EI [cm$^{-6}$kpc$^3$]\footnote{Emission Integral (EI$=\int$n$^2$dV) is calculated from the best-fitted value of the normalization factor of the APEC component of spectra  CGM, using the equation: $norm=\frac{10^{-14}}{4\pi D^2}\times$EI, where D is the comoving distance to NGC\,3221} &  $\chi^2/dof$  \\
\hline
off1 & 0.115$^{+0.023}_{-0.014}$ & 1.105$^{+0.359}_{-0.323}$ & 179.78/195 \\
off2 & 0.121$^{+0.034}_{-0.022}$ & 0.617$^{+0.313}_{-0.359}$ & 206.30/194 \\
\decimals
\hline
\end{tabular}
\end{table}

The above analysis shows that we can determine the EI of the CGM of NGC\,3221 with $3.4\sigma$ significance, once detected. However, it does not tell us whether the signal from the galaxy's CGM is required in the spectral model. To assess the same, we performed an F-test with models excluding/including the CGM component. We obtained an F statistic value $=15.3525$, for $196$ dof, with a null-hypothesis probability of $0.00012$ when fitted against the off-field1. This confirms the detection of the CGM of NGC\,3221 with 99.988\% confidence (this is statistical significance only; as  discussed above, there are also systematic uncertainties in determining the foreground/background). 
\subsection{Confirmation as the extragalactic signal}
\noindent While the detection of a signal in the galaxy-field is confirmed, the poor spectral resolution of XIS makes it difficult to distinguish between a local and an extragalactic emission. Therefore, to further confirm that the detection is truly from the CGM of NGC\,3221, we consider the following possibilities. \\
1) The emission cannot be from the 0.525 keV \oi line, as that is already included in the spectral model, and is contributing to the 0.575 keV emission folded through the instrumental profiles. Also, the \oi emission peaking at $\approx$ 0.575 keV would imply a huge blueshift ($v\sim0.1c$), which is extremely unlikely. \\
2) The geocoronal SWCX has a characteristic \ovii emission at 0.575 keV. However, the geocoronal emission depends on the solar wind proton flux \citep{Henley2015b}. As the average proton flux during the galaxy-field observation is lower than the proton fluxes during the off-field observations (figure \ref{fig:SWCX}), the geocoronal SWCX is expected to be lower in the galaxy-field than the off-fields. So, the excess emission in the galaxy-field is unlikely to be associated with geocoronal SWCX. Rather, assuming similar foreground in the galaxy-field and the off-fields makes our detection conservative.          \\
3) The temperature of the detected signal in the galaxy-field is comparable with that of the LHB. Therefore, the signal may be attributed to the variation of the LHB emission. However, as the observed fields are separated by $\sim$2 $\deg$, this would imply a density and/or path length fluctuation of LHB at a length scale of 3--7 pc. Such small scale inhomogeneity and anisotropy of the LHB has not been observed \citep{Kuntz2000,Snowden2014,Liu2017}. So, the detected emission is possibly not from the LHB, although we cannot rule it out completely.    \\
4) The temperature of the detected signal in the galaxy-field (10$^{6.1}$ K) is consistent with the temperature of the MWH measured from \textit{ROSAT} maps \citep{Snowden2000,Kuntz2000}, and is well within the temperature distribution of the MWH \citep{Henley2013}. Therefore, we cannot rule out the possibility that the detected signal is from the halo of Milky Way, and not from NGC\,3221. However, the temperature of the MWH in our off-fields and the galaxy-field (10$^{6.3-6.4}$ K) are higher than the temperature of the detected signal. Therefore, if the detected signal is from the MW CGM, it would imply that the galaxy-field has two distinct temperature components of the MWH, while the 2 nearby off-fields have only temperature component of the MWH; which is highly unlikely.

From the above discussion we can conclude, both physically and statistically, that the detected signal is highly likely to be extragalactic, from the CGM of NGC\,3221.

\begin{figure}
\begin{subfigure}
\centering
\includegraphics[trim=8 0 5 25,clip,scale=0.475]{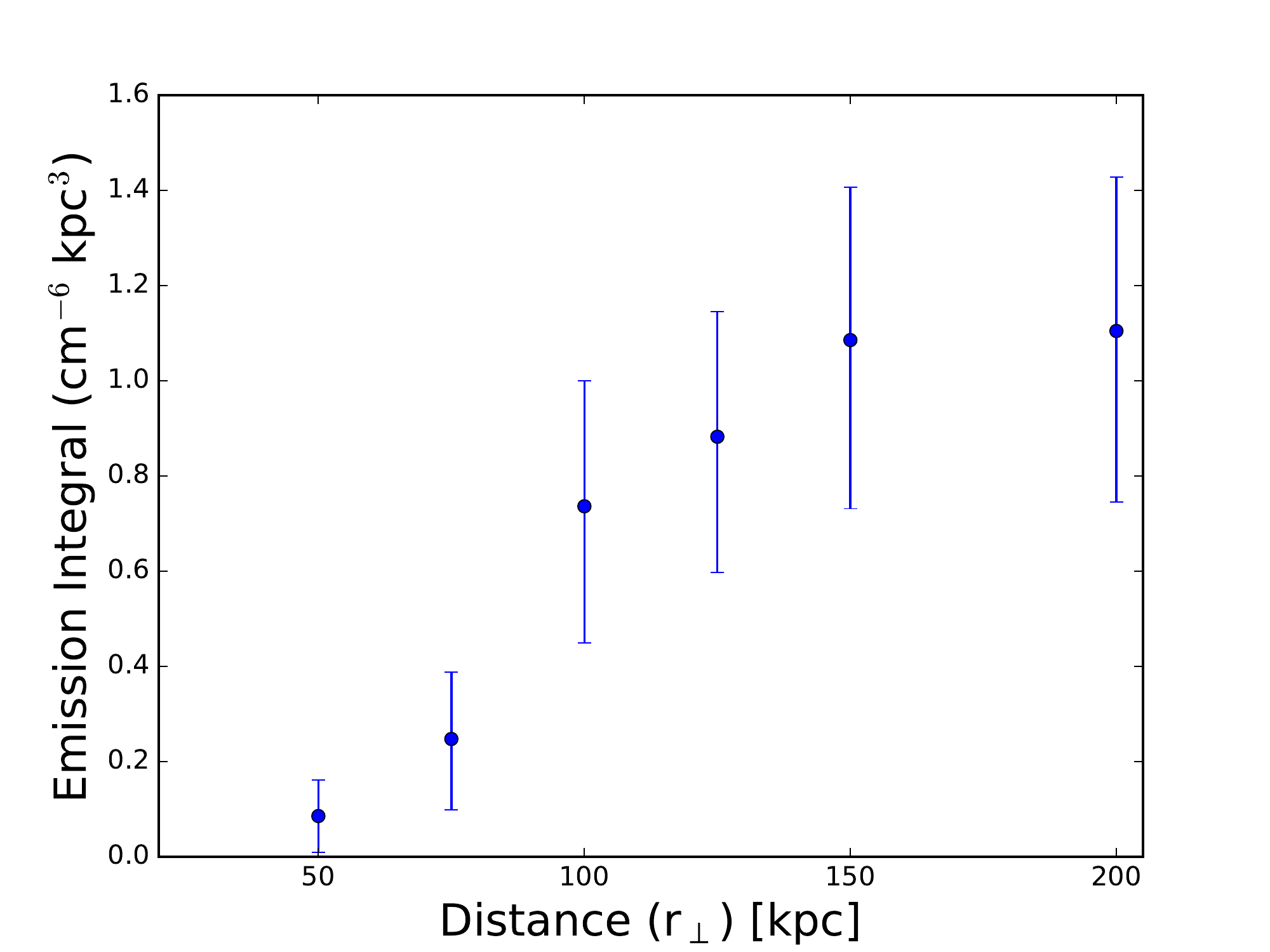}
\end{subfigure}
\caption{The radial profile of the emission integral (EI) \textcolor{black}{of the warm-hot CGM of NGC\,3221}} \label{fig:EI_profile}
\end{figure}

Excited by the \textcolor{black}{first ever detection} of the CGM around NGC\,3221, we now determine the radial profile of the EI. For this purpose, we select several annular regions of 25 kpc inner radius (transverse distance $r_\perp$) and 50, 75, 100, 125, 150, and 200 kpc outer radius in the galaxy-field. In the off-fields, we select the corresponding regions of similar geometric area and perform similar spectral analysis as discussed above (\S\ref{sec:spec_analysis}) for the whole field. For each annular region in the galaxy field, we use a corresponding annular region in the off-field to model the background/foreground. We allow each component of the spectra to vary independently in every annular region, so that the spatial and spectral variation of each component are taken into account. The average temperature in all of the annular regions are found to be the same within statistical uncertainties with log($T_{avg}/K) = 6.1\pm0.1$, indicating an isothermal medium. As the galaxy is not at the center of the field (figure \ref{fig:soft_src}, left panel), we cannot extract complete annuli beyond 100kpc (r$_\perp$) from the center of the galaxy, and miss a fraction of the CGM emission. We correct for the area of those segments, assuming an isotropic/azimuthally symmetric emission from the CGM of NGC\,3221, to obtain the cumulative emission integral profile over the spherical volume around NGC\,3221. In figure \ref{fig:EI_profile} we show the radial profile of the EI. \textcolor{black}{At smaller radii, it increases with radius, and then flattens. A local source like LHB and/or MWH would have almost uniform surface brightness over the galaxy-field. Hence, their emission integral will be proportional to radius squared, overshooting the saturated profile observed in figure \ref{fig:EI_profile}. This reinstates that the detected signal is from the CGM of NGC\,3221 with a radially falling surface brightness (as also seen in figure \ref{fig:surf-rad}), and eliminates the possibility of it being a Galactic emission.} As the emission integral is a cumulative measurement, its saturation indicates how far the warm-hot diffuse medium is extended. We find that the radial profile of emission integral becomes flat beyond 150 kpc from the center of NGC\,3221, giving a rough estimate of the spatial extent of the CGM of the galaxy down to our sensitivity limit. 
%----------------------------------------------------
\subsection{Modeling}\label{sec:modelling}
\begin{figure}
\centering
\includegraphics[trim=30 0 10 7,clip,scale=0.5]{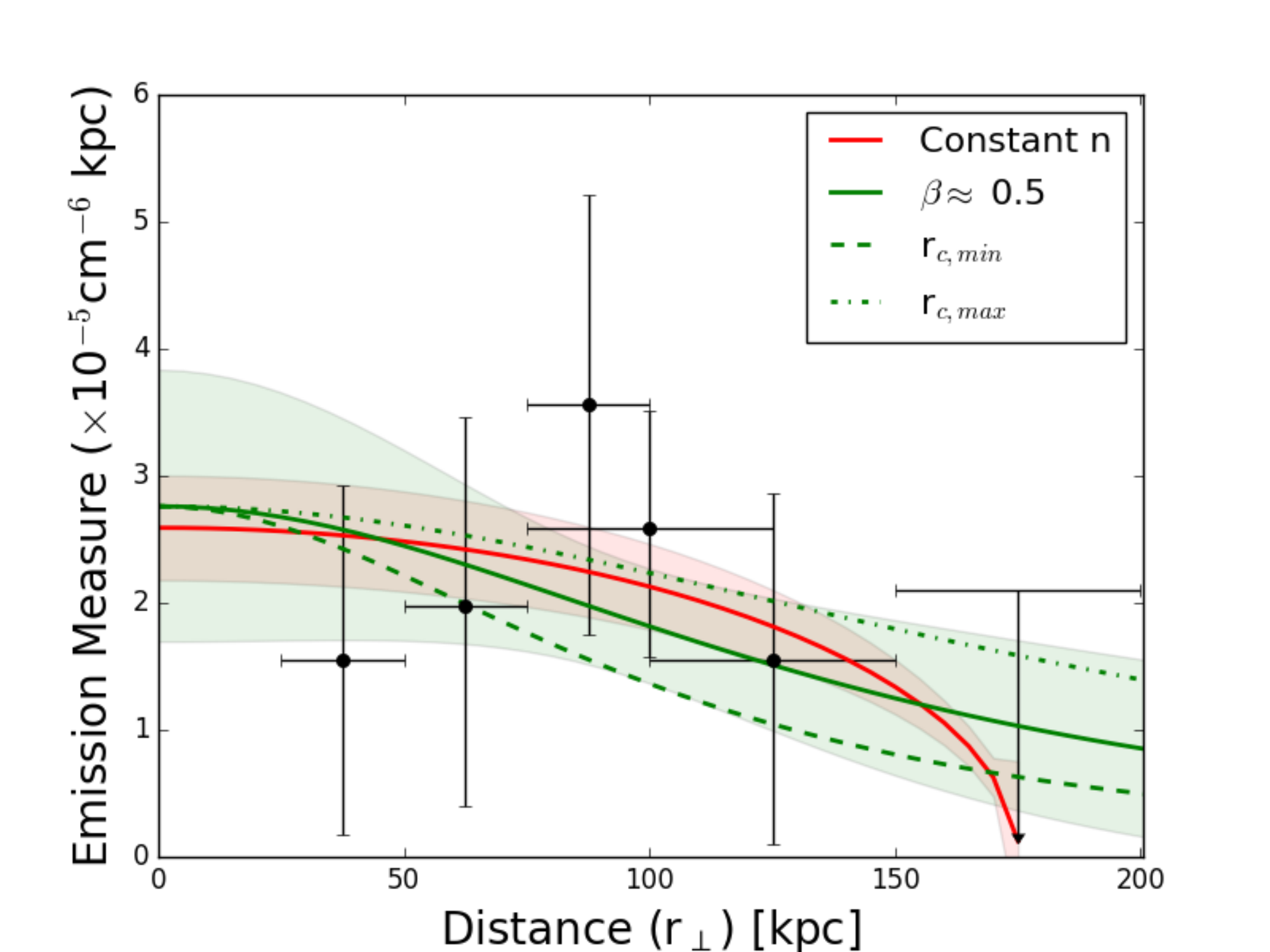}
\caption{Radial profile of emission measure (EM). The best-fitted
  constant density and $\beta$-model with $\beta \sim$ 0.5 are shown, with shaded regions describing 1$\sigma$ ranges. As
  we do not have any measurement at the center, the uncertainty at r=0
  is quite high. The allowed range of $r_c$ ([110,225] kpc for model A)
  for the best-fitted values of $EM_o$ and
  $\beta$ has been calculated empirically } \label{fig:EM-fit}
\end{figure}

%-----------------------------------------------------------
\noindent Our next goal is to estimate the density, spatial extent and the total
baryonic mass contained in the CGM of NGC\,3221, {\textcolor{black}{given that the detected signal is truly from NGC\,3221}}. We calculate the average emission measure (EM) profile from the cumulative emission integral profile. As the emission integral of the larger annulus encompasses the emission integral of the smaller annulus, the uncertainties are not independent. We assume that the correlation coefficient of the consecutive measurements is 0.5, i.e. the covariance is half the geometric mean of their own variance: $\sigma_{i,i-1}= 0.5\sigma_i\sigma_{i-1}$. We fit the EM profile with a $\beta-$model.  The $\beta$-model (eqn. \ref{eq:n_beta}) has a density profile:
\begin{equation}\label{eq:n_beta}
n = n_o(1 + (r_\perp/r_c)^2 )^{-\frac{3\beta}{2} }
\end{equation}
where $n_o$ is the central density and $r_c$ is the core radius,
resulting in an EM profile similar to the surface brightness profile
(eqn. \ref{eq:EM_beta}) \citep{Sarazin1986},
\begin{equation}\label{eq:EM_beta}
EM = EM_o (1 + (r_\perp/r_c)^2 )^{-3\beta+0.5}
\end{equation}
where $EM_o$ is the central emission measure.
Because of the small number of data points, large error bars and the
degeneracy between the parameters, we could not fit the three parameters
of the $\beta$-model simultaneously. Therefore, as we did for the
surface brightness profile in \S\ref{sec:image_analysis}, we fit the EM profile in two different ways. \\ 
1) A $\beta-$model: We first fixed $\beta=0.5$ and fit the radial profile for
EM$_o$ and r$_c$. Then we fix r$_c$ to the best-fit value (152.4 kpc)
and fit for EM$_o$ and $\beta$. The resulting fit ($\chi^2/dof=3.3/4$)
yielded parameters values: EM$_o$ = $(2.8\pm1.1) \times 10^{-5}$
cm$^{-6}$ kpc and $\beta= 0.56\pm 0.36$. We could not determine the range
of r$_c$ from the fit, but we obtain the range of $r_c$ empirically for
the best-fitted values of $EM_o$ and $\beta$ which is consistent with
the 1$\sigma$ error span of the emission measure profile, with minimum
$r_c=110$ kpc and maximum $r_c=225$ kpc. Interestingly, the best-fitted values and the confidence intervals are consistent with those obtained by fitting the radial profile of
surface brightness in \S\ref{sec:image_analysis}.  Additionally, we fit the EM profile with Model B, a constant density model. \\
2) A constant density model: The truncated constant density homogeneous medium (n = $n_o$(constant)),
has an elliptical profile of projected EM (eqn. \ref{eq:EM_homo}),
\begin{equation}\label{eq:EM_homo}
EM = 2n_o^2R_{out}\sqrt{1-(r_\perp/R_{out})^2}
\end{equation}
where R$_{out}$ is the spatial extent of the gaseous medium.  We obtain
best-fitted ($\chi^2$/dof = 1.4/4) values of n = $\sqrt{n_in_e}$= (2.7
$\pm$ 0.2) $\times$ 10$^{-4}$cm$^{-3}$ and R$_{out}$= 175 $\pm$ 2 kpc
(figure \ref{fig:EM-fit}).
%------------------------------------------
\subsection{Physical characterization}\label{sec:phys_charac}
%------------------------------------------
\noindent We calculate the mass of the gaseous halo and the central
electron density assuming $n_e = 1.3n_i$ and the mean atomic mass
$\mu=0.62$, the estimates for fully ionized gas of solar
metallicity. The mass in the constant density model (Model B) measured
out to $R_{out}=175$kpc is $(17 \pm 1) \times 10^{10}$ M$_\odot$
and $n_e$ = 3 $\times $10$^{-4}$cm$^{-3}$. For the
$\beta$-model (Model A), we calculate the central electron
density using eqn \ref{eq:n_o}
\begin{equation}\label{eq:n_o}
n_{eo}n_{io} = \frac{EM_o}{ 2\int_{0}^{R_{out}} (1 + (r/r_c)^2 )^{-3\beta} dr }
\end{equation}
and the mass using eqn. \ref{eq:Mass},
\begin{equation}\label{eq:Mass}
M =  4\pi(n_{eo}+n_{io})\mu m_H\int_{0}^{R_{out}} (1 + (r/r_c)^2 )^{-\frac{3\beta}{2}} r^2 dr 
\end{equation}
and quote the values in table \ref{tab:Mass-beta-A}. All these mass values are quoted for solar metallicity. Since the halo metallicity is likely sub-solar \citep[e.g.][]{Gupta2012,Prochaska2017}, and the estimated halo mass is inversely related to metallicity, these values are lower limits. We confidently detect the CGM emission out to 150 kpc, so the first outer radius is set to $R_{out}=150$ kpc. The constant density model has a cutoff radius of $175$ kpc, so we calculate mass within 175 kpc as well. The \textit{Suzaku} field of view extends to 200 kpc, so we calculate the mass and density parameters $R_{out}=200$ kpc as well, even though the photon count rate within 200 kpc and 150 kpc is small. The region beyond 200 kpc is outside the field of view of \textit{Suzaku}; therefore to calculate the CGM mass within the virial radius of the galaxy (calculated below), we extrapolate the $\beta$-profile out to the virial radius $R_{200}= 253$ kpc (see Table \ref{tab:Mass-beta-A}). 
\begin{table}[t]
\renewcommand{\thetable}{\arabic{table}}
\centering
\caption{Mass and density estimates (assuming Z = Z$_\odot$) of the warm-hot CGM for the $\beta$ model} \label{tab:Mass-beta-A}
\begin{tabular}{c c c}
\tablewidth{0pt}
\hline
\hline
R$_{out}$[kpc] & Mass[10$^{10}$ M$_\odot$] & $n_{eo}$[10$^{-4}$cm$^{-3}$] \\
\hline
150 & 10.16$\pm$1.69 &  4.16$\pm$0.69\\
175 & 14.221$\pm$2.248 &  4.03$\pm$0.64\\
200 & 18.868$\pm$2.903  & 3.94$\pm$0.60   \\
253 & 31.38$^{+9.17}_{-7.80}$  & 3.80$\pm$0.59   \\
\hline
\end{tabular} 
\end{table} 

We compare the mass estimates of the two models out to 150 kpc. In the $\beta-$model, the mass is $(10\pm2) \times 10^{10}$ M$_\odot$ and in the constant density model it is $(11\pm 1)\times 10^{10}$ M$_\odot$. Thus the mass estimates in the two models are consistent with each other, showing that the mass measurement is convergent. 

We estimate the virial mass ($\simeq M_{200}$) of NGC\,3221 from the
maximum rotational velocity of the galaxy ($V_{max}$) using the baryonic
Tully-Fisher relation for cold dark matter cosmology. We take the average of 14 measurements of $V_{max}$ from HyperLeda catalog\footnote{\url{http://leda.univ-lyon1.fr/}
\citep{Makarov2014}}, which gives $V_{max} = 269 \pm 18$ km
s$^{-1}$. We calculate $M_{200}$ using
eqn. \ref{eq:M_vir} from the numerical results of
\citet{Navarro1997},
\begin{equation}\label{eq:M_vir}
log(\frac{M_{200}}{10^{10}M_\odot}) = -5.31 + (3.23\pm0.03)log(V_{max})
\end{equation} 
and obtain $M_{200}$= (3.44$\pm$0.94)$\times$10$^{12}$ M$_\odot$. We
compute the virial radius ($\simeq R_{200}$) from M$_{200}$ using
eqn. \ref{eq:R_vir},
\begin{equation}\label{eq:R_vir}
M_{200} = 200\times \frac{4\pi}{3}\rho_{crit} R_{200}^3
\end{equation} 
where $\rho_{crit}$ is the critical density of the universe. We thus
estimate $R_{200}$ = 253$\pm$23 kpc. 
The virial temperature is:
\begin{equation}
    T_{vir}= \frac{GM_{200}m_{eqv}}{3k_BR_{200}};\hfill m_{eqv}=\mu(m_p+m_e)
\end{equation}
We find that $T_{vir}= 1.5 \pm 0.4 \times 10^6$ K is comparable with the average
temperature of the warm-hot CGM, showing that the CGM can be in thermal and hydrostatic equilibrium.

We further compare the predicted value of $\beta$ with the value we obtain from
the data. For an isothermal hydrostatic gas, \citet{Sarazin1986} predicts that the $\beta$ index is linked to the ratio of gravitational and thermal energy density in the form of
eqn. \ref{eq:beta_theo},
\begin{equation}\label{eq:beta_theo}
\beta_{predict} = \mu m_H \sigma_v^2/3k_BT_X
\end{equation}
where $\sigma_v$ and $T_X$ are the velocity dispersion of the galaxy and
the average temperature of the warm-hot halo gas respectively. Neglecting the
difference between $V_c$ and $V_{max}$, we calculate $\sigma_v$ from \cite{Corsini2004}:
\begin{equation}\label{eq:v_disp}
V_c = (1.32\pm0.09)\sigma_v + (46\pm14).
\end{equation}
We use $T_X = T_{avg}$ and $\mu$=0.62 as reported and discussed in
\S\ref{sec:detect} and \S\ref{sec:phys_charac} respectively. There are 14 studies available in HyperLeda catalog giving a wide range of values of $V_{max}$, from $V_{max}$= 234.9$\pm$0.8 km/s \citep{Wong2006} to 291.05$\pm$8.0 km/s \citep{Schneider1990}, with a mean of 268.65$\pm$18.0 km/s. Using these values, we obtain a range of $\beta_{predict}$ from $0.4\pm 0.1$ to $0.7\pm0.1$, with a central value of $\beta_{predict}=0.6\pm 0.1$. This is similar to the best-fit
$\beta=0.56\pm 0.36$ obtained for the $\beta-$model radial density profile in \S\ref{sec:modelling}. 
Thus the warm-hot CGM in NGC\,3221 appears to be a isothermal hydrostatic system.
%---------------------------------------
\section{Discussion}\label{sec:discuss}
%--------------------------------
\noindent \textcolor{black}{Assuming that our detection of the  emission from the warm-hot CGM of NGC\,3221 is true, we have determined the physical parameters of the CGM such as its temperature, density profile and mass, with a varying degree of uncertainty (\S\ref{sec:results}). Similarly}, in the following subsections we use the measured and derived parameters to draw interesting inferences on the missing feedback, missing baryons, and missing metals problems. 
%--------------------------------------------
\subsection{Missing feedback}
%------------------------------------
\noindent \citet{Wang2010} showed that the X-ray luminosity of hot gas in galaxies is only a few percent of the energy injected by supernovae; this is called the ``missing feedback'' problem. With our \textcolor{black}{detection} of a large amount of warm-hot gas in the CGM of NGC\,3221 we investigate the role of starburst-driven winds and supernovae on the physical characteristics of the CGM and their relevance to the missing feedback problem.
%---------------------------------
\subsubsection{The role of starburst-driven winds}
%----------------------------------------------------
\noindent \textcolor{black}{The metals in the CGM come from the stellar winds and supernova feedback. Because we have detected Oxygen in the warm-hot CGM of NGC\,3221, it is necessary to check if the stellar feedback of the galaxy can actually enrich the CGM with metals.} According to the standard superbubble theory of disk galaxies
\citep{Maclow1988}, the blowout of the gas from the galaxies is
determined by the energy injection rate per unit disc area. Following
\citet{Henley2010}, we calculate the SN rate from $R_{SN} =$ 0.2
L$_{\small{FIR}}$/10$^{11}$L$_\odot$ and obtain the specific SN rate
$F_{SN} = R_{SN}/(0.25\pi D_{25}^2)$ \citep{Strickland2004} to be
$\approx 88$ Myr$^{-1}$kpc$^{-2}$. This value is much larger than the
critical surface rate for superbubble blowout of $F_{SN} >$ 25-40
Myr$^{-1}$kpc$^{-2}$ \citep{Maclow1988}, indicating that the warm-hot
X-ray corona can be potentially enriched by the starburst-driven winds,
leading to non-zero metallicity of the gaseous medium. 
%----------------------------------------------------
\subsubsection{Energy budget of the galactic corona}
%-----------------------------------------------------
\noindent For the total energy output from supernovae (SNe), we assuming 10$^{51}$ erg per SN.  The explosion rate of Type Ia SNe is calculated using \cite{Mannucci2005}:
\begin{equation}
\nu_{Ia} = 0.044 (M_\star/ 10^{10} M_\odot)\ {\rm century}^{-1}
\end{equation}
and that of core-collapse (CC) SNe is calculated using \cite{Heckman1990}:
\begin{equation}
\nu_{CC} = 0.77 (SFR/ M_\odot yr^{-1})\ {\rm century}^{-1}
\end{equation}
Using the value of M$_\star$ and SFR from table \ref{tab:galaxy}, we obtain $\dot{E}_{SN}$ = $\dot{E}_{Ia}$+$\dot{E}_{CC}= (1.4\pm0.2) + (24\pm3) \times 10^{41}$ erg s$^{-1}$. The X-ray (0.1-2.0 keV) luminosity $L_X$ within 150 kpc (the extent out to which significant emission is detected) is $\approx 5.1^{+1.7}_{-1.5} \times$10$^{41}$ erg s$^{-1}$. Thus, the X-ray radiation efficiency is $\eta=
L_X/\dot{E}_{SN}\approx$ $0.2 \pm 0.1$. This indicates that a significant fraction ($\approx 10-30$\%) of SNe energy has been converted into soft X-ray emission, alleviating the missing feedback problem. The
rest of the energy would be in the hot gas beyond 150 kpc, in the hot gas that cooled, and in the mechanical energy of the outflow.
%-----------------------------------
\subsection{Missing Baryons}\label{sec:missing_baryon}
%---------------------------------
\noindent In order to determine the baryon census of NGC\,3221, \textcolor{black}{we consider the warm-hot CGM mass (modulo metallicity), the mass of other phases of CGM, the mass of stellar and ISM components, and the virial mass.}  
%-----------------------------
\subsubsection{Baryonic mass}
%---------------------------------
\noindent We first consider the mass in the warm-hot phase that we detect with \suzaku. In table \ref{tab:Mass-beta-A} we have presented the
mass for solar metallicity. However, the mass depends (inversely) on the metallicity: 
\begin{equation}
    M(Z) = M\times\Big(\frac{Z}{Z_\odot}\Big)^{-0.5}
\end{equation}
%, which is shown in figure \ref{fig:M_hotCGM}.  
%\begin{figure}
%\centering
%\includegraphics[trim=12 0 45 30,clip,scale=0.45]{Mass_hot_Z_onemodel.eps}
%\caption{Mass of warm-hot CGM at different radius for a range of metallicity. The shaded regions show the $1\sigma$ range} \label{fig:M_hotCGM}
%\end{figure} 
%We have secure detection of the CGM emission out to 150 kpc and upper limit out to 200 kpc; the models are extrapolated beyond this radius. The figure shows that 
The metallicity is a significant unknown in the derived mass of the warm-hot CGM. The differences in the predicted mass due to different density profiles are negligible compared to the differences caused by the (unknown) metallicity. The warm-hot CGM mass for $\frac{1}{3}$ $Z_\odot$ at a conservative radius of 150 kpc is $M_{hot,halo} = 15.8\pm2.6 \times 10^{10}$ \msun (table \ref{tab:Mass-missing}). This shows that the warm-hot CGM of NGC\,3221 harbors a huge amount of gas. 
\begin{table}[t]
\renewcommand{\thetable}{\arabic{table}}
\centering
\caption{Mass budget of NGC\,3221} \label{tab:Mass-missing}
\begin{tabular}{c c }
\tablewidth{0pt}
\hline
\hline
Phase & Mass (\msun)\\
\hline
M$_\star$ & 10.00$\pm$1.26 $\times$10$^{10}$\\
M$_{\hi}$& 1.64$\pm$0.20 $\times$10$^{10}$\footnote{Using the value of $D_L$ from table\ref{tab:galaxy}, and combining the statistical uncertainty and the calibration error of (0.11$\pm$0.16)$\times$10$^{10}$ M$_\odot$} \\ 
M$_{H_2}$ & 8.59 $\times$10$^8$\footnote{Corrected using the updated value of CO-to-H$_2$ conversion factor from \citet{Yao2003}} \\
M$_{dust,ISM}$ & 1.41$\pm$0.15 $\times$10$^8$\\
\hline
M$_{disk}$ & 11.74$\pm$1.28 $\times$10$^{10}$\\
\hline
M$_{dust,halo}$ & 5.00 $\times$10$^7$\footnote{Obtained from \citet{Menard2010}}\\
M$_{cold,halo}$ & 3.80$\pm$0.72 $\times$10$^8$\footnote{Calculated for Z = Z$_\odot$ using the median value of M$_{\caii,halo}$ for high-mass star-forming galaxies from \citet{Zhu2013}} \\
M$_{cool,halo}$ & 9.20$\pm$4.30 $\times$10$^{10}$\footnote{{Taken from \citet{Prochaska2017}} for the gas out to 160 kpc}\\
M$_{warm,halo}$ & 4.53$\pm$0.75 $\times$10$^{10}$ \footnote{Calculated using the ratio of warm and hot gas from the fiducial model of \citet{Faerman2017} and the mass of the hot gas we obtain in our work. As the temperature of the CGM of the galaxies in their model is similar to ours, we assume that the relative abundance of the warm and the hot gas remains same}\\
%M$_{warm,halo}$(B) & 5.01$\pm$1.54 $\times$10$^{10}$ \\
M$_{hot,halo}$ & 15.77$\pm$2.62 $\times$10$^{10}$\footnote{The mass within 150 kpc at Z= $\frac{1}{3}$ $Z_\odot$, the median metallicity of cool CGM at low redshift \citep{Prochaska2017}}  \\
%M$_{hot,halo}$(B) & 17.44$\pm$5.38 $\times$10$^{10}$ \\
\hline
M$_{b,halo}$ & 29.54$\pm$5.09 $\times$10$^{10}$\\
%M$_{b,halo}$(B) & 31.70$\pm$7.06 $\times$10$^{10}$\\
\hline
\hline
M$_{b,tot}$ & 41.28$\pm$5.25 $\times$10$^{10}$\\
%M$_{b,tot}$(B) & 43.44$\pm$7.17 $\times$10$^{10}$ \\
M$_{vir}$ & 3.44$\pm$0.94\textcolor{black}{$^{+1.87}_{-1.58}$} $\times$10$^{12}$\footnote{Hyperleda catalog \citep{Makarov2014}}\\
\decimals
\hline
\hline
\end{tabular} 
\end{table}

In table \ref{tab:Mass-missing}, we have included the cold (T $<10^4$K), cool (T $\approx 10^{4-5}$K) and warm (T $\approx 10^{5-6}$K) phases of the CGM, obtained from literature for L$^\star$ galaxies. We find that the warm-hot gas is the most dominant mass component accounting for $53\pm9$\% of the CGM mass. 

Next, we compare the warm-hot gas mass to the disk mass (M$_{disk}$). We get M(\hin) and M(H$_2$) from \citet{Thomas2002},
and calculate the interstellar dust mass using eqn. \ref{eq:dust} from \citet{Peeples2014},
\begin{equation}\label{eq:dust}
log(M_{dust}/M_\odot) =  0.86 log(M_\star/M_\odot) - 1.31
\end{equation}
By adding all these components to the stellar mass, we obtain M$_{disk}$ = (M$_\star$ + M(\hin) + M(H$_2$) + M$_{dust}$) $\simeq$ $12\pm1$ $\times$ 10$^{10}$ M$_\odot$. We see that the mass in the warm-hot CGM is more than that in the galactic disk; this is similar to what was found for the Milky Way \citep{Gupta2012,Nicastro2016b}. 
%----------------------------------------------------
\subsubsection{Baryon fraction}
%--------------------------------------------------------
\noindent It is of interest to know whether the baryons in the warm-hot CGM can account for the missing baryons of NGC\,3221. We calculate the baryon fraction $f_b= \frac{M_{b,tot}}{M_{vir}}$ and compare that to the the cosmological baryon fraction $f_{b,cosmo} = 0.157\pm0.001$ \citep{Planck2016}. We calculate the total baryonic mass of NGC\,3221 (M$_{b,tot}$) by adding the mass in the disk (M$_{disk}$) to the mass in the halo (M$_{b,halo}$).  However, the calculation of $f_b$ is complicated by the fact that the virial mass of NGC\,3221 is highly uncertain, with different studies providing different values of the rotational velocity (see \S\ref{sec:phys_charac}). This leads to a factor of 2 dispersion in the estimated virial mass, \textcolor{black}{and similar systematic uncertainty in the baryon fraction. The baryon fraction of the hot CGM is $f_{b,hot} = \frac{M_{hot,halo}}{M_{vir}}$ = 0.080 $\pm$ 0.023 $^{+0.068}_{-0.032}$ (Z/$\frac{1}{3}$Z$_\odot$)$^{-0.5}$, showing that it is the reservoir of at least 50\% baryons of the baryon budget of NGC\,3221. The total baryon fraction is $f_b$ = 0.120 $\pm$ 0.036 $^{+0.104}_{-0.048}$.} This clearly closes the baryon budget of NGC\,3221. 

\textcolor{black}{The uncertainty in the baryon fraction is three-fold. First, the statistical uncertainty of the detected signal is propagated to the density profile and hence, the mass of the warm-hot CGM. Second, the metallicity adds a factor of few systematic uncertainty to the mass of the warm-hot CGM.} The uncertainty in the virial mass, however, is the major source of uncertainty in $f_b$.
%-----------------------------------------------------
\subsection{Missing metals}
%-----------------------------------------------------
\noindent In addition to the missing baryons problem discussed above, there appears to be a missing metals problem
\citep{Peeples2014}. The total mass of metals produced in the universe appears to be much larger than that found in galaxies. In our discovery of the warm-hot gas in the CGM of NGC\,3221, what we actually detect is Oxygen, so we are in a good position to determine whether metals in the CGM alleviate the missing metals problem. 

The total mass of the metals expected to be produced in a galaxy with the stellar mass of NGC\,3221 is M$_{metal,expect}$ = 6.31 $\pm$ 0.80 $\times$ 10$^9$ M$_\odot$, and the metals found in stars, ISM and interstellar dust is only M$_{metal,disk} \approx$ 1.99 $\times$ 10$^9$ M$_\odot$ \citep{Peeples2014}. We calculate the mass of metals in the X-ray emitting warm-hot phase as a function of metallicity from the mass of the warm-hot halo gas within 150 kpc. We assume the bulk mass fraction of metals in the Sun to be Z$_{frac}$ = 0.0142 \citep{Asplund2009}. The metal fraction is then $f_{metal} = \frac{M_{metal,found}}{M_{metal,expected}}$, where $M_{metal,found} = M_{metal,disk} + M_{metal,CGM}$. The metal fraction within 150 kpc is 0.45$^{+0.13}_{-0.10}$ for $\frac{1}{3}$Z$_\odot$; and it ranges from about 0.30 to 0.71, increasing slowly with metallicity from 0.1 to 1.0 Z$_\odot$. Thus we see that about 30-70\% of the metals are still missing \textcolor{black}{within the observed volume of the CGM.} This suggests that a large fraction of the galactic metals got expelled \textcolor{black}{at a larger distance, within and beyond the CGM.} This is consistent with earlier observations \citep{Peeples2014} and the results from the 3D, high-resolution hydrodynamic simulations \citep{LiMiao2017}, showing that metals are preferentially expelled from a galaxy. 
%------------------------------------------------
\section{Comparison with earlier observations}\label{sec:compare}
%--------------------------------------------------
\noindent \citet{Gupta2012} combined their column density measurements of z=0 \ovii and \oviii absorption lines with the emission measure of MW halo from literature, and found a massive warm-hot (T $\approx$ 10$^{6.1-6.4}$K) CGM extended out to 239$\pm$100 kpc. The mass calculated for solar abundance was 1.2$\times$10$^{10}$ M$_\odot$, being comparable with the baryonic mass of the disk $\approx$ 6$\times$10$^{10}$ M$_\odot$. They showed that the baryonic fraction $f_b$ of this warm-hot gas varies from 0.09 to 0.23 depending on the estimates of M$_{vir}$= (1.0-2.5)$\times$10$^{12}$ M$_\odot$, bracketing the universal value of $f_b$. This was later confirmed by \citet{Nicastro2016b}, using independent observations.   

However, the picture is quite different around other MW-type spiral galaxies. Using \chandra observations, \citet{Strickland2004} found diffuse X-ray emitting halos around eight nearby (D$<$ 17 Mpc) galaxies extending only out to $\approx$ 10 kpc. With \textit{XMM-Newton}, \citet{Tullmann2006} searched for the soft X-ray emission around 9 nearby star-forming galaxies, and did not find any diffuse emission beyond 4-10 kpc around the disk, after correcting for projection. \citet{Yamasaki2009} using \textit{Suzaku} observation detected the X-ray halo of NGC4631 (D$\approx$ 8 Mpc) extending out to about 10 kpc from the galactic disk. \citet{Bogdan2015} searched for hot gaseous coronae around 8 nearby (14 $<$ D $<$ 40 Mpc) normal (SFR $<$10 M$_\odot$yr$^{-1}$) spiral galaxies with M$_\star$ $\approx (0.7-2.0) \times$ 10$^{11}$M$_\odot$ using \chandra observations \citep[see also][]{Li2013}. They did not detect any statistically significant diffuse X-ray emission beyond the optical radii ($\approx$ 20 kpc) of the galaxies. Thus all the detections were of extra-planar gas, not of the extended CGM. \textcolor{black}{The non-detections were caused by a variety of reasons: either the observations were too shallow (t$_{exp} \approx$ 8-55 ks), or the choice of the detectors was not optimal (i.e. they had small field-of-view and/or small effective area), or the targets were so close that only a small fraction of the halo was within the field-of-view, or the targets were so far that the emission was too faint to be detected in the observed time. Additionally, in the spectral analyses, the sky backgrounds were often extracted from the same fields of the targets, usually from regions within the virial radii. Thus, if the CGM is extended out to the virial radii, the backgrounds would be overestimated and the CGM signal would be underestimated. Therefore, the non-detections of the CGM in other MW-like external galaxies are likely the effect of observational designs and/or analysis methods, rather than the intrinsic properties of the galaxies themselves.} 

On the other hand, searches for the CGM emission around individual and stacked massive spirals yielded detections \citep{Anderson2011,Dai2012,Bogdan2013a,Anderson2016,Li2017,Bogdan2017,Li2018,Bregman2018}. The masses of hot extended gaseous halos were found to be huge, but the virial masses were too large, so the total baryon fraction was small, with $f_b \le 0.1$.

Thus, we can summarize the previous work on X-ray observations of the CGM as follows: (1) in the MW, a large mass of warm-hot gas is detected that can close the baryon census; (2) in other MW-type galaxies, the mass of the hot gas, if detected, is not significant, though this is likely an observational bias; and (3) in massive galaxies, a large mass of hot gas is detected, but it is insufficient to account for the missing baryons. \textcolor{black}{\citet{Tullmann2006} observed NGC\,3221 with \xmmn. Since they did not detect the CGM of this galaxy while we did with our \suzaku observations, it is imperative to understand why. In their imaging analysis, \citet{Tullmann2006} performed contour mapping of the soft X-ray image of the galaxy,  including the disk. The surface brightness of the disk and the foreground (local/Galactic) sources are much higher than that of the halo gas, therefore the contours were not sensitive to the faint extended emission. In our imaging analysis, we removed the inner 25 kpc region around the galaxy (\S\ref{sec:data-process}; figure \ref{fig:soft_src}, left panel), and focused on the region beyond that. This ensured that the X-ray emission we study is not contaminated by the galaxy itself. With our quantitative imaging analysis we find an indirect evidence of the halo gas (\S\ref{sec:image_analysis}, figure \ref{fig:surf-rad}). Additionally, we performed a detailed and intricate spectral analysis (\S\ref{sec:spec_analysis}), where the galaxy-field is simultaneously fitted with two blank sky observations to estimate the foreground, background and the atmospheric contamination (figure \ref{fig:spectral-fit}).} This was crucial in extracting the weak CGM signal, but there is no spectral analysis in \citet{Tullmann2006}. To conclude, we \textcolor{black}{detected} the CGM of NGC\,3221, but \citet{Tullmann2006} did not, because of a combination of differences in imaging and spectral analysis.

Ours is the first discovery of the warm-hot CGM of an L$^\star$ galaxy, with a 3.4$\sigma$ significance, extended out to at least 150 kpc from the center of the galaxy. For a metallicity of $\frac{1}{3}$Z$_\odot$, the warm-hot gas accounts for 53$\pm$9\% of the CGM mass and 38$\pm$5\% of the total baryonic mass of the galaxy, being the dominant component of all the baryons. It is the first external spiral galaxy accounting for all its missing galactic baryons.
%------------------------------------------------------
\section{Conclusion and Future directions}
%------------------------------------------------------
\noindent In this paper we have studied the warm-hot X-ray-emitting CGM of an
L$^\star$ star-forming spiral galaxy NGC\,3221, with deep
\textit{Suzaku} observations. We rediscovered the \oi contamination in
\suzaku\, data using a robust spectral modeling taking into account all
foreground and background components. The
contaminating \oi emission line needs to be included in the spectral
modeling, otherwise it can result in false detection and/or
overestimation of the \ovii signal and wrong characterization of the
CGM, as in \cite{Kataoka2013,Miller2016}. Our science results
are as follows:\\
1) We have \textcolor{black}{detected} the warm-hot CGM around NGC\,3221 at 
  3.4$\sigma$ significance out to at least 150 kpc. The bolometric X-ray
  luminosity of the gas is $\approx 5 \times 10^{41}$ erg s$^{-1}$. This
  is the first detection of an extended CGM around an external L$^\star$
  spiral galaxy. \\
2) The main uncertainty in estimating the baryon fraction is the
  virial mass of the the galaxy, but we find that $f_b$ can be
  consistent with the cosmic value. This is the first external spiral
  galaxy with all the baryons accounted for.  It shows that the warm-hot
  phase of the CGM of L$^\star$ spirals can account for the missing
  baryons.\\
3) A significant fraction ($\approx 20$\%) of the supernova energy
  has been converted in soft X-ray emission, alleviating the missing
  feedback problem.\\
4) Our study of the metal fraction in the CGM indicates that the
  metals are preferentially expelled from the galaxy. 

To search for and characterize the warm-hot CGM further, it is essential to study a broad sample of galaxies with a range of M$_\star$, SFR and M$_{vir}$. At present, \textit{XMM-Newton} is the most suitable mission to detect the faint emission from the warm-hot halo gas because of its large effective area and large FOV. On a longer timescale, upcoming missions (e.g. \textit{XRISM,Athena,Lynx}) in the next decade and beyond will offer an outstanding opportunity to observe the warm-hot diffuse medium in absorption in unprecedented detail. This will bring us closer to understanding galaxy formation and evolution. 
%--------------------------------------------------------------------
\section*{Acknowledgements}
\noindent We gratefully acknowledge support through the NASA ADAP grants NNX16AF49G to SM and 80NSSC18K0419 to AG. Support for this work was also provided by the National Aeronautics and Space Administration through Chandra Award Number 25-10-20-30300-53001 to AG issued by the Chandra X-ray Observatory Center, which is operated by the Smithsonian Astrophysical Observatory for and on behalf of the National Aeronautics Space Administration under contract NAS8-03060. YK acknowledges support from the grant PAIIPIT IN106518.
\facilities{\textit{Suzaku}} %, AAVSO, CTIO:1.3m, CTIO:1.5m,CXO}
\software{HeaSoft v6.17 \citep{Drake2005}}, AstroPy v1.1.2 \citep{Astropy2013}, NumPy v1.11.0 \citep{Dubois1996}, SciPy v0.17.0, Matplotlib v1.5.3 \citep{Hunter2007} 
%--------------------------------------------------------------------
%--------------------------------------------------------------------------
\bibliographystyle{aasjournal}
%\bibliography{reference}

%---------------------------------------------------------------------------
\appendix

\section{Data reduction of \textit{Suzaku}}\label{data-reduction}

\begin{figure}[h]
\begin{subfigure}
\centering
\includegraphics[trim= 0 5 55 35,clip,scale=0.33]{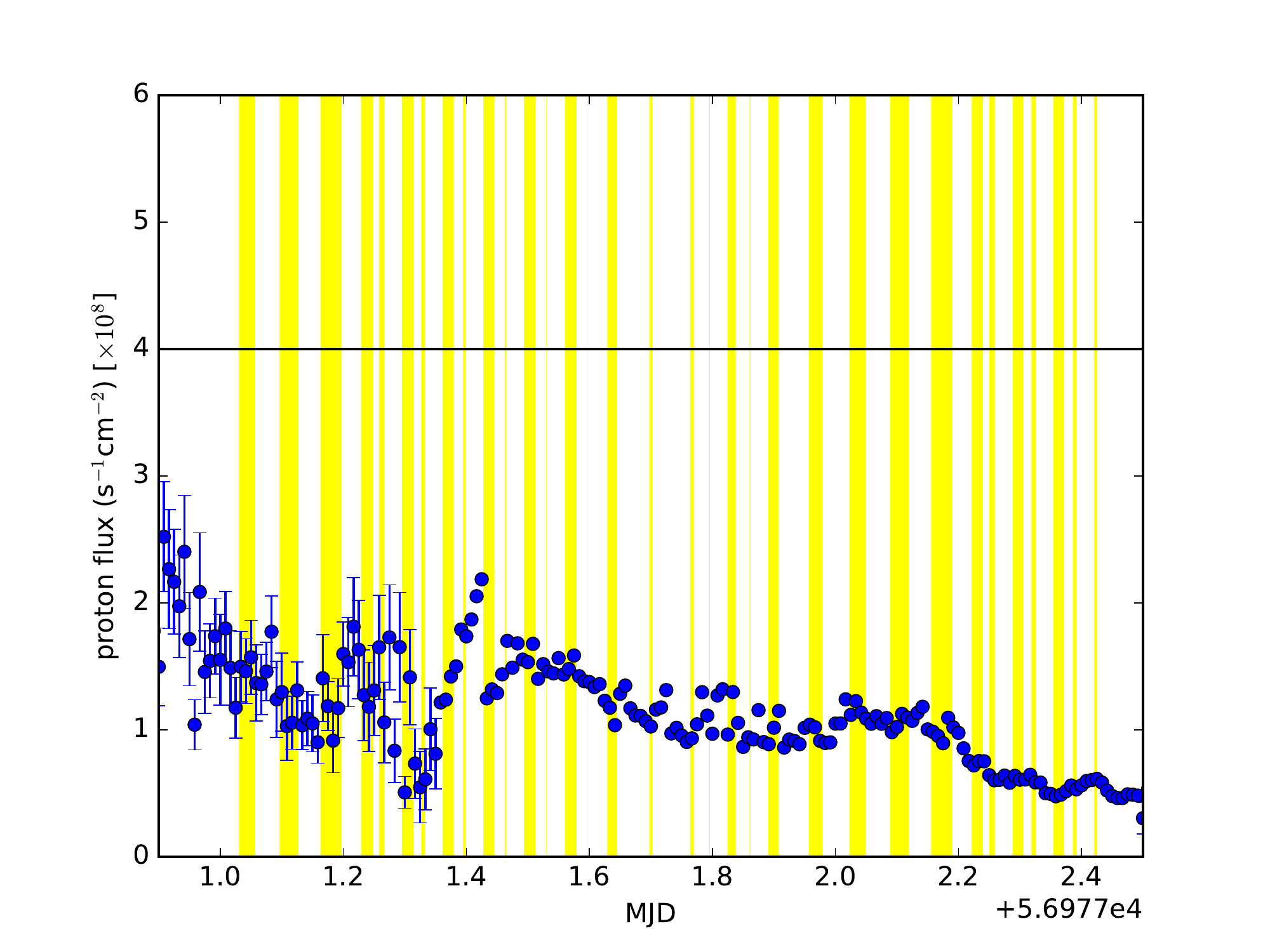}
\end{subfigure}
\begin{subfigure}
\centering
\includegraphics[trim=0 5 55 35,clip,scale=0.33]{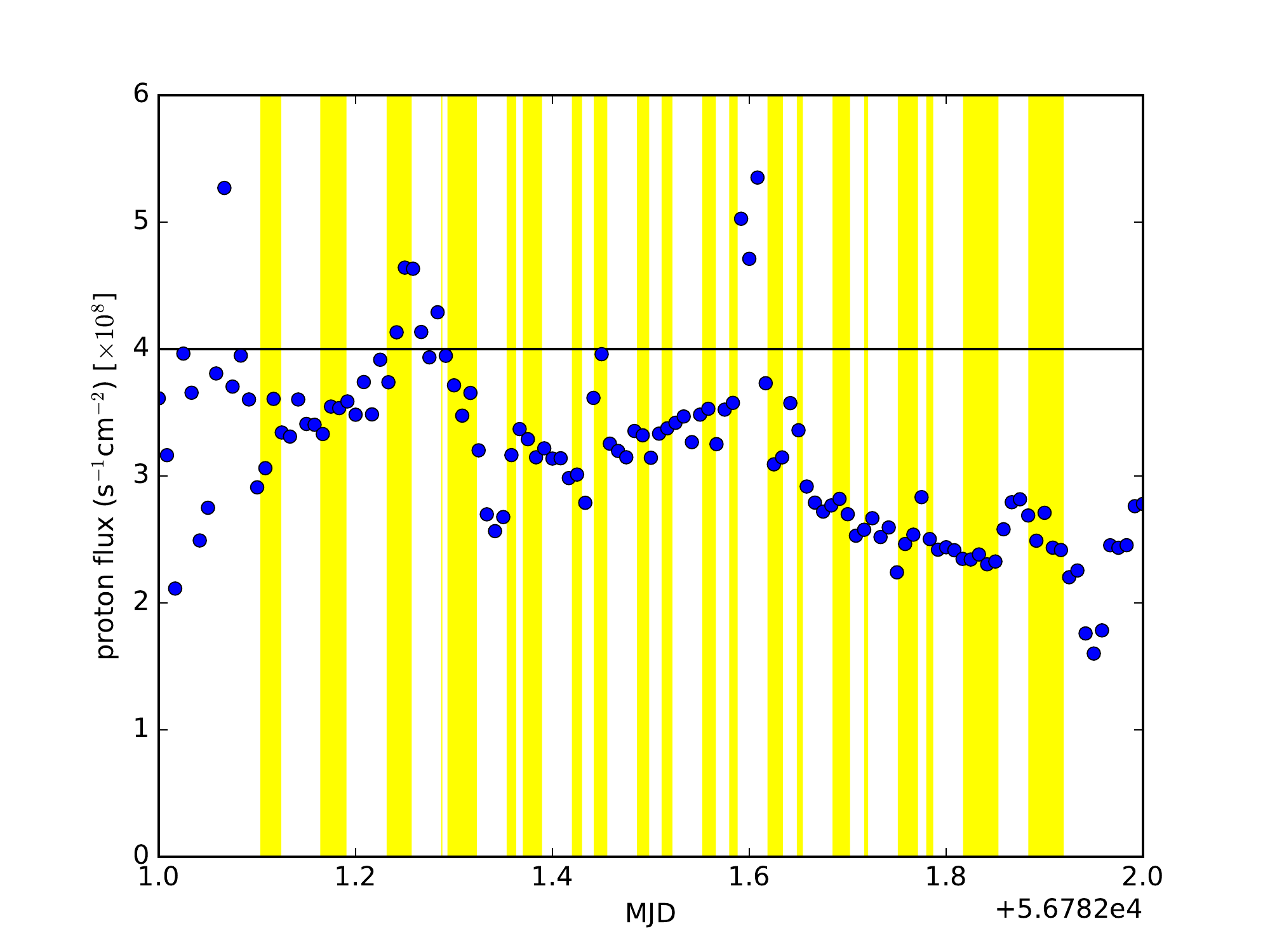}
\end{subfigure}
\begin{subfigure}
\centering
\includegraphics[trim=0 5 55 35,clip,scale=0.33]{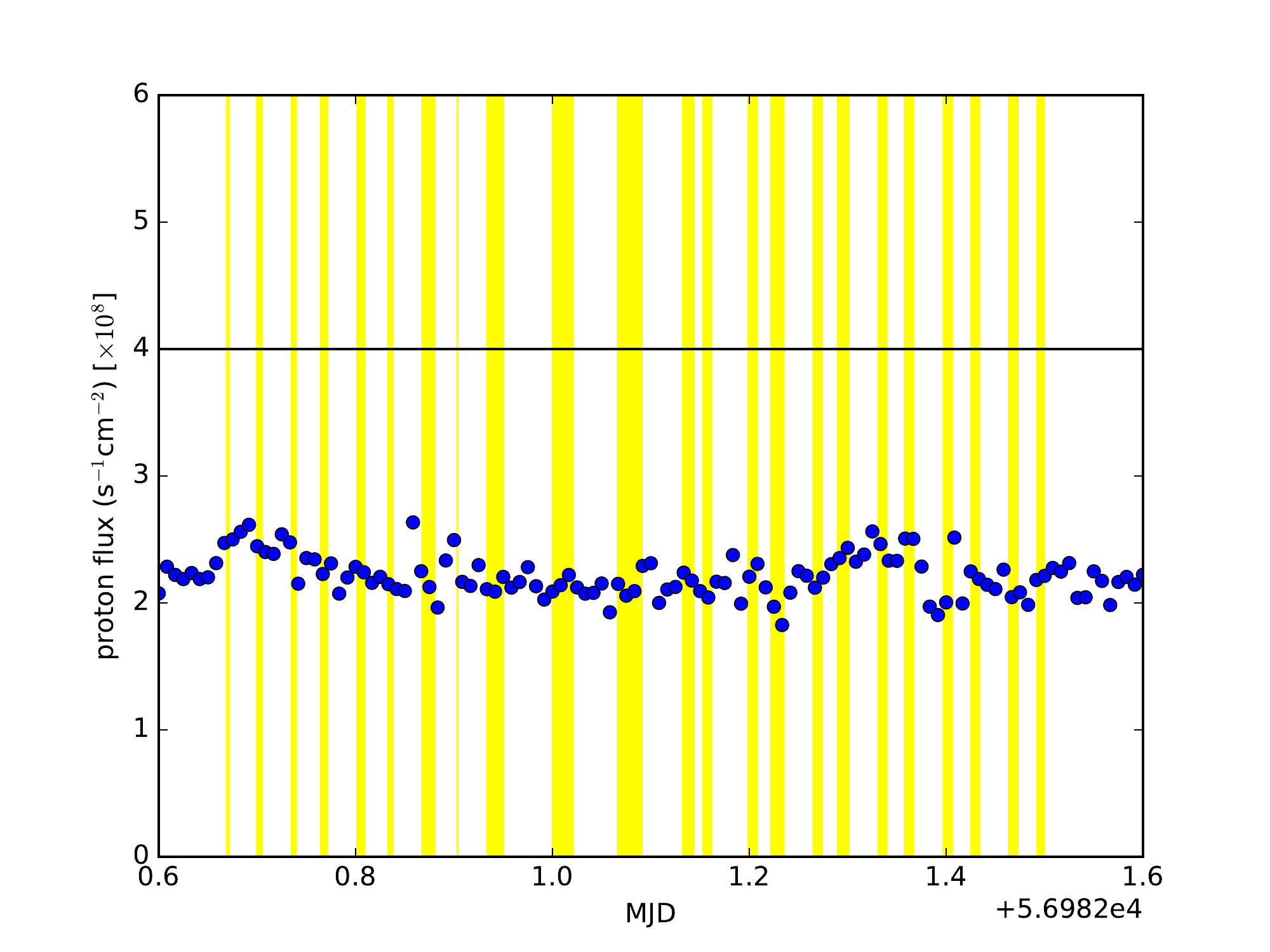}
\end{subfigure}
\caption{Solar wind proton flux during observations. In the middle and
  right panel the error bars are smaller than the size of data
  points. The shaded regions are GTIs of respective
  observations. Left: galaxy-field, middle and right:
  off-fields \label{fig:SWCX}}
\end{figure}

\noindent We have worked with unfiltered event files and reprocessed
everything using the updated calibration database. As we do not follow
the default conditions of
Aepipeline\footnote{\url{https://heasarc.gsfc.nasa.gov/docs/suzaku/analysis/abc/}}
in every step, the data reduction procedure is sketched in brief :
\begin{enumerate}
\item Using \texttt{ftmerge} we combine the components of 3x3 and 5x5
  mode observations separately.
\item We update the time and coordinate information using {\tt{xistime}}
  and {\tt{xiscoord}} respectively. We correct the Euler angles in the
  attitude file using \texttt{aeattcor}, which calculates the effects of
  the ``thermal wobbling" caused by thermal distortions of the satellite
  bodies. To have an accurate mapping of channel to energy, we update
  the one-to-one relation between channel no. and energy using
  \texttt{xispi}. We also update the information of pixel quality using
  {\tt{xisputpixelquality}} with the latest bad column file. We create
  the good time intervals (GTI) for each of the 3x3 and 5x5 mode
  observations using \texttt{xisgtigen}.
\item In {\tt{xselect}} we apply GTI correction; remove hot and
  flickering pixels using {\tt{sisclean}} and screen some events based
  on their grades. We remove the noisy pixels at higher bit, the pixels
  illuminated by the $^{55}$Fe calibration sources, and the events in
  the second trailing rows of artificial charge injection at 6keV based
  on their status information. The effective exposure time after GTI
  correction reduces to 41 ks for galaxy-field, 30 ks for off-field1,
  and 23 ks for off-field2.  

  We remove the events recorded during or within 436 seconds after the
  passage through South Atlantic Anomaly (SAA). We allow an elevation
  angle (ELV) of $>10 \deg$ and the elevation angle from the day Earth
  rim (DYE$\_$ELV) of $>20\deg$ to reduce the contamination from Earth's
  atmosphere.\footnote{\textcolor{black}{While a larger cut-off value of DYE$\_$ELV (e.g. $>40\deg$ or $>60\deg$) removes practically all atmospheric contamination as shown by \cite{Sekiya2014}, the signal-to-noise ratio becomes too low to fit a multi-component spectrum. Therefore, we continue with DYE$\_$ELV $>20\deg$, and model the atmospheric contribution as a spectral component.}} Also, we impose a strict condition on geomagnetic cut-off
  rigidity COR2 $>$8 to obtain a good enough signal-to-noise ratio by
  reducing the particle background. It must be clarified that the older
  version of COR2, COR cannot do the filtering as efficiently as COR2
  and hence, the filtered data can be substantially different in the
  energy range of interest if the updated version is not used.  \\
  
  We find that the proton flux in the solar wind\footnote{Data is taken
    from \url{http://www.srl.caltech.edu/ACE/ASC/}} never exceeds the
  typical threshold of $2.0-4.0\times10^8$ cm$^{-2}$s$^{-1}$ during the GTI
  (figure \ref{fig:SWCX}), making sure that effect of the geocoronal SWCX
  (solar wind charge exchange) induced emission is small and stable.

\item We merge the screened 3x3 and 5x5 mode data in \texttt{xselect} to
  obtain the full exposure.
\end{enumerate}

Then, we remove the point sources from all fields and remove the
galaxy's contribution from the galaxy-field before extracting the
spectra (\S\ref{sec:data-process}).\\

Using {\tt{xisnxbgen}} we extract the non X-ray background spectrum
(NXB) of each annular region with the latest NXB event file and bad
column file. We screen out the events in the second trailing rows of
artificial charge injection at 6keV based on its pixel quality.\\

We extract the redistribution matrix function (RMF) of the whole region
using {\tt{xisrmfgen}}. As the current {\tt{xisrmfgen}} does not
consider spatial variation of spectral response on the CCD chip, which
is negligible for the current data, we do not calculate RMF for each
annular region separately.\\

We generate the ancillary response files (ARF) of each annular region
using {\tt{xissimarfgen}} assuming a uniform source of $20'$ radius. We
include the updated bad column file and remove the second trailing rows
of artificial charge injection at 6keV based on its pixel quality.
%-----------------------------------------------------------------
\section{Modeling the emission measure profile with free $\beta$}
\noindent In the $\beta$-model mentioned in the main text, we forced $\beta$ to be around 0.5, as is assumed for galaxy clusters and the halos of massive galaxies. We can perform a more rigorous analysis by allowing $\beta$ to be free. We fit for $EM_o$ and $r_c$ by varying $\beta$ in the range of 0.1-1.0 in steps of 0.01, and find that the best fit ($\chi^2/dof \approx 3.945/4$) is obtained for $\beta=0.19$, with the best-fitted $r_c= 22.19$ kpc. Then, we fit for $EM_o$ and $\beta$
keeping $r_c$ fixed at $22.19$ kpc and obtain (with $\chi^2/dof$ =
3.662/4) $EM_o= (3.2 \pm 2.6)\times 10^{-5}$ cm$^{-6}$ kpc and $\beta =
0.24\pm0.10$. We determine the range of $r_c$ empirically
for the best-fitted values of $EM_o$ and $\beta$ which is consistent
with the 1$\sigma$ error span of the emission measure profile, with minimum $r_c=10$ kpc and maximum $r_c=45$
kpc. Interestingly, the best-fitted values and the confidence intervals
are consistent with those obtained by fitting the radial profile of
surface brightness in \S\ref{sec:image_analysis} for both Model A and the free-$\beta$ model. As expected, a steeper $\beta$ requires a larger core radius
$r_c$ (Model A) and a flatter $\beta$ has a smaller $r_c$ (free-$\beta$ model). We find that the mass and the central density of the CGM in the two models (tables \ref{tab:Mass-beta-A} and \ref{tab:Mass-beta-B}) are consistent within 1$\sigma$. So, with the current data we cannot prefer one model over the other, although $\chi_\nu^2$ for the model with variable $\beta$ is closer to 1 than that for the model with constrained $\beta$.  
\begin{figure}[h]
\centering
\includegraphics[trim=30 0 10 7,clip,scale=0.5]{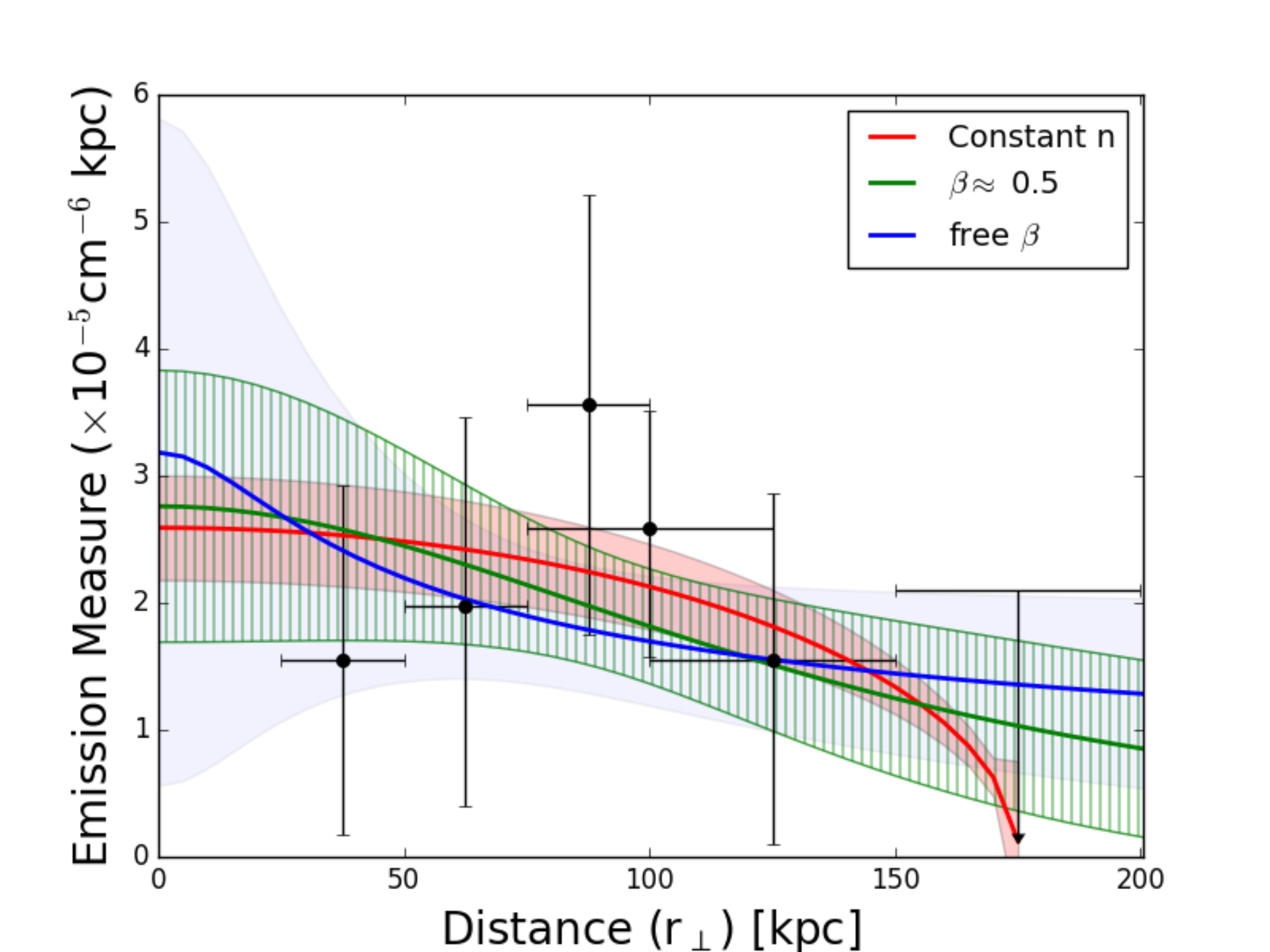}
\caption{Radial profile of emission measure (EM). The best-fitted
  constant density and $\beta$-models with $\beta \sim$ 0.5 and free
  $\beta$ are shown, with shaded and hatched regions describing 1$\sigma$ ranges. As
  we do not have any measurement at the center, the uncertainty at r=0
  is quite high.} \label{fig:EM-fit_three}
\end{figure}
\begin{table}
\renewcommand{\thetable}{\arabic{table}}
\centering
\caption{Mass and density estimates (assuming Z = Z$_\odot$) of the warm-hot CGM for the free-$\beta$ model.}\label{tab:Mass-beta-B}
\begin{tabular}{c c c}
\tablewidth{0pt}
\hline
\hline
R$_{out}$[kpc] & Mass[10$^{10}$ M$_\odot$] & $n_{eo}$[10$^{-4}$cm$^{-3}$] \\
\hline
%150 & 10.16$\pm$1.69 &  4.16$\pm$0.69\\
%257 & 31.38$^{+9.17}_{-7.80}$  & 3.80$\pm$0.59   \\
150 & 7.977$\pm$2.571 &  6.90$\pm$2.22\\
175 & 11.242$\pm$3.465 &  6.78$\pm$2.09\\
200 & 15.135$\pm$4.486  & 6.58$\pm$1.85   \\
253 & 26.465$^{+12.015}_{-9.971}$  & 6.54$^{+1.88}_{-1.80}$   \\
\hline
\end{tabular} 
\end{table} 
%----------------------------------------------------------------------
\section{Assumptions, Caveats and uncertainties}
\noindent We calculate the propagation of uncertainties using the truncated series expansion of error:
\begin{equation}
    \sigma_y^2 = \bigg(\frac{\partial f}{\partial a}\bigg)^2 \sigma_a^2 + \bigg(\frac{\partial f}{\partial b}\bigg)^2\sigma_b^2 + 2\bigg|\frac{\partial f}{\partial a}\bigg| \bigg|\frac{\partial f}{\partial b}\bigg| \sigma_{ab}
\end{equation}where $y= f(a,b)$. The variance of (and the covariance between) the fitting parameters of the emission measure profile (central emission measure $EM_o$, core radius $r_c$ and the slope $\beta$) have been used to calculate the uncertainties of (most of) the quantities in \S\ref{sec:phys_charac} and \S\ref{sec:discuss}. As the parameters are strongly correlated (e.g. steeper profile has larger core radius), the uncertainties of the functions are not necessarily of the same order of the uncertainties of the parameters. We should also clarify that the emission measure has been calculated from the cumulative emission integral (EI) as opposed to the differential (bin-by-bin) emission integral. So, the derived errors of the emission measure (figure \ref{fig:EM-fit}) are subject to the assumption of a moderate covariance between the adjacent annuli.  As the surface brightness falls off radially from the galaxy (figure \ref{fig:surf-rad}), the photon count in the annuli (those we used in \S\ref{sec:image_analysis}) are not sufficient to extract the CGM's contribution from the multi-component spectrum dominated by the foreground/background. This led us to redesign the annuli size (as discussed in \S\ref{sec:modelling}) and calculate the cumulative EI (figure \ref{fig:EI_profile}).       \\   

We have calculated the mass of the warm-hot halo for the best-fitted value of $r_c$ assuming full ionization, and with  volume-filling factor ($f_v$) and covering factor ($f_c$) of 1. We do not include the errors in $r_c$ to calculate the uncertainties in the central density, total mass and $f_b$. As $r_c$ is poorly constrained, inclusion of its error reduces the confidence in determining the baryon fraction $f_b$, giving us no extra information. The dependence of mass on $f_{ion}=\frac{n_e}{n_i}$ at a given ionization state is very weak, $\frac{(f_{ion}+1)}{\sqrt{f_{ion}}}$ decreasing/increasing by a factor of $\approx$1.01 from Z=Z$_\odot$ to Z=0/5Z$_\odot$, respectively. The value of mean atomic mass at a given metallicity is smallest at full ionization, so  our calculated value puts the lower limit of the mass of the warm-hot gas. \textcolor{black}{The uncertainty due to absolute metallicity (Metal/H) discussed in \S\ref{sec:missing_baryon} is based upon the assumption that O/H is a proxy for Metal/H, such that the abundance ratio of Oxygen and other metals are same as solar. Because our data are not sensitive to different chemical compositions, we opt for the solar value, which is the simplest.} Also, we have assumed that the ions and the electrons follow the same $\beta$-profile. As we find the warm-hot CGM to be isothermal out to 150 kpc, and there is no information about the metallicity gradient, taking $\frac{n_e}{n_i}$ = radially constant remains a valid assumption. We obtain the number density from the projected emission measure, where any directional property is averaged out. Also, while calculating the emission measure from the emission integral, we assumed that the CGM is isotropic, because we found that any angular dependence of the azimuthal profile of surface brightness out to 100kpc does not affect the radial profile (figure \ref{fig:surf-az}). \\

The absorption study of the internal structure of the CGM \citep{Koyamada2017} suggests that the high ionization states (e.g. \ovi ion) are found in larger clouds and  in the diffuse medium than the low ionization states (e.g. \ciii ion). Extrapolating this for \ovii, we assume that the volume-filling factor and covering factor are unlikely to be considerably smaller than 1. Without any measurement of the anisotropy and clumping we cannot estimate $f_c$, and thus calculate the mass for $f_c = 1$ and $f_v = 1$ only. This is consistent with the semi-analytic models of \citet{Faerman2017} as well. \\

Our used values of M$_\star$ and $D_L$ are larger by 4\% and $\approx$8\% than some other studies in literature, respectively. However, using the smaller values of these quantities we found that the revised value of $f_b$ with 1$\sigma$ error bar still brackets the universal value of $f_b$.  
\end{document}